\documentclass{aa}
\usepackage{txfonts}
\usepackage{graphicx}
\newcommand{\bm}[1]{\mbox{{\boldmath $#1$}}}

\newcommand{\pder}[3]{\frac{{\partial}^{#3} {#1}}{{\partial} {#2}^{#3}}}
\newcommand{\pc}{\mbox{pc} \ }
\newcommand{\paral}{\parallel }
\newcommand{\kpc}{\mbox{kpc} \ }

\newcommand{\Myr}{\mbox{Myr} \ }

\newcommand{\cm}{\mbox{cm} \ }

\newcommand{\km}{\mbox{km} \ }
\newcommand{\s}{\mbox{s}}
\newcommand{\erg}{\mbox{erg}}

\newcommand{\GeV}{\mbox{GeV}}

\newcommand{\muG}{\mu{\mbox{G}}}
\newcommand{\Rm}{{\mbox{Rm}}}
\begin{document}
\title{Cosmic ray driven dynamo in galactic disks.}

\subtitle{A parameter study}

\author{Micha\l{} Hanasz\inst{1}, Katarzyna Otmianowska-Mazur\inst{2},
        Grzegorz Kowal\inst{2,3}, Harald Lesch\inst{4}}
\institute{Toru\'n Centre for Astronomy, Nicolaus Copernicus University,
           PL-87-148 Toru\'n/Piwnice, Poland
      \and Astronomical Observatory, Jagiellonian University,
           ul. Orla 171, 30-244 Krak\'ow
      \and Department of Physics and Astronomy, McMaster University,
           1280 Main St. W., Hamilton, ON L8S 4M1, Canada
      \and Astronomical Observatory, Munich University,
           Scheinerstr. 1, D-81679, Germany}

\offprints{M. Hanasz}

\date{Received ...; accepted ...}

\abstract{}
{We present a parameter study of the magnetohydrodynamical dynamo driven by
cosmic rays in the interstellar medium (ISM) focusing on the efficiency of
magnetic field amplification and the issue of energy equipartition between
magnetic, kinetic and cosmic ray (CR) energies.}
{We perform numerical CR-MHD simulations of the ISM using the extended version of
ZEUS-3D code in the shearing box approximation and taking into account the
presence of Ohmic resistivity, tidal forces and vertical disk gravity. CRs are
supplied in randomly distributed supernova (SN) remnants and are described by
the diffusion-advection equation, which incorporates an anisotropic diffusion
tensor.}
{The azimuthal magnetic flux and total magnetic energy are amplified in majority
of models depending on a particular choice of model parameters. We find that the
most favorable conditions for magnetic field amplification correspond to
magnetic diffusivity of the order of $3\times 10^{25} \cm^2\s^{-1}$, SN rates
close to those observed in the Milky Way, periodic SN activity corresponding to
spiral arms, and highly anisotropic and field-aligned CR diffusion. The
rate of magnetic field amplification is relatively insensitive to the magnitude
of SN rates in a rage of spanning 10\% up to 100\% of realistic values. The
timescale of magnetic field amplification in the most favorable conditions is
150 Myr, at galactocentric radius equal to 5 kpc, which is close to the timescale of galactic rotation. The final
magnetic field energies reached in the efficient amplification cases fluctuate
near equipartition with the gas kinetic energy. In all models CR energy exceeds
the equipartition values by a least an order of magnitude, in contrary to the
commonly expected equipartition. We suggest that the excess of cosmic rays in
numerical models can be attributed to the fact that
the shearing-box does not permit cosmic rays to leave the system along the
horizontal magnetic field, as it may be the case of real galaxies.}
{}


\maketitle

%

\section{Introduction}

An attractive idea of {\em fast galactic dynamo} has been proposed by
Parker (1992). The idea relies on two ingredients: (1) cosmic rays (CR)
continuously supplied to the disk by supernova (SN) remnants and (2) fast
magnetic reconnection which operates in current sheets and allows to dissipate
and relax the random magnetic field components in the limit of vanishing
resistivity. Over the last decade we have investigated the different elements,
physical properties and consequences of Parker's idea and scenario by means of
analytical calculations and numerical simulations (Hanasz \& Lesch 1993,
1997, 1998, 2000, 2001, 2003a, 2003b, Hanasz et al. 2002, 2004, 2006,
Lesch \& Hanasz 2003; Otmianowska-Mazur et al. 2003, 2007;
Kowal et al. 2003, 2005)

The first complete 3D numerical model of the CR-driven dynamo has been
demonstrated by Hanasz et al. (2004, 2006). In this paper we perform a parameter
study of the CR-driven dynamo model by examining the dependence of magnetic
field amplification on magnetic diffusivity, supernova rate determining the CR
injection rate, temporal modulation of SN activity, grid resolution, and CR
diffusion coefficients.

The principle of action of the CR-driven dynamo is based on the cosmic ray
energy supplied continuously by SN remnants. Due to the anisotropic diffusion of
cosmic rays and the horizontal magnetic field configuration, cosmic rays tend to
accumulate within the disc volume. However, the configuration stratified by
vertical gravity is unstable with respect to the Parker instability.  Buoyancy effects
induce vertical and horizontal motions of the fluid and formation of undulated
patterns -- magnetic loops in frozen-in, predominantly horizontal magnetic
fields. The presence of rotation in galactic disks implies a coherent twisting
of the loops by means of the Coriolis force, which leads to the generation of
small-scale radial magnetic field components. The next phase is merging of
small-scale loops by the magnetic reconnection process to form the large scale
radial magnetic fields. Finally, the differential rotation stretches the radial
magnetic field to amplify the large-scale azimuthal magnetic field component.
The coupling of amplification processes of radial and azimuthal magnetic field
components results in an exponential growth of the large scale magnetic field.
The timescale of magnetic field amplification, resulting from the action of the
CR-driven dynamo, has been found (Hanasz et al. 2004, 2006) to be equal to 140-
250 Myr, depending on the galactocentric radius, which is comparable to the
galactic rotation period.

The CR-dynamo experiments reported in the afore mentioned papers rely on 
energy of CRs accelerated in SN remnants.  Recently Gressel et al. (2008a,b)
reported a series of non-ideal MHD simulations demonstrating dynamo action
resulting from the SN-driven turbulence, in absence of CRs. Using a similar set
of galactic disk parameters, with angular velocity enlarged 4 times the  value
of $0.025 \Myr^{-1}$, typical for the galactocentric radius of Sun, these
authors found amplification of the large scale magnetic fields on a time scale
of $250 \Myr$. This indicates some similarity between the CR-driven dynamo and
the dynamo driven by thermal energy output from supernovae. The similarity is
presumably related to the buoyancy effect which can be commonly attributed to
the excess of both the thermal and cosmic ray energies in the disk volume.

The magnitudes of galactic magnetic fields are usually estimated from
measurements of the radio synchrotron emission arising from the acceleration of
cosmic ray electrons in the magnetic field. To interpret the radio emission
spectrum, it is usually assumed that the energy density in the magnetic field is
of the same order of magnitude as the energy density in cosmic ray protons
(which are assumed to outnumber the electrons by 100:1, as they do in our
Galaxy). There is however no compelling evidence of energy equipartition. Since
the equipartition or minimum energy assumption is one of the few ways to 
calculate radio source parameters, it is important to determine how
reasonable the approach generally is. Recently Strong et al. (2007) and Snodin
et al. (2005) raised again the question about the applicability of the
equipartition argument.

From the observational point of view (Fitt and Alexander 1993; Vallee 1995) the
equipartition assumption seems to hold. In particular, Vallee's comparison of
three different methods determining galactic magnetic field strengths (Faraday
rotation method, equipartition method and Cosmic ray equipartition) shows that
the equipartition fields are in a quite good agreement. On the other hand, Beck
and Krause (2005) considered in detail a problem which was raised by Chi and
Wolfendale (1993), that the commonly used classical equipartition or
minimum-energy estimate of total magnetic field strengths from radio synchrotron
intensities is of limited practical use because it is based on the hardly known
ratio K of the total energies of cosmic ray protons and electrons and also has
inherent problems. They present a revised formula using the {number density
ratio} {K} for which they give estimates. For particle acceleration in strong
shocks K is about 40 and increases with decreasing shock strength. Their revised
estimate for the field strength gives larger values than the classical estimate
for flat radio spectra with spectral indices of about 0.5-0.6, but smaller
values for steep spectra and total fields stronger than about 10 $\muG$. In very
young supernova remnants, for example, the classical estimate may be too large
by up to 10. On the other hand, if energy losses of cosmic ray electrons are
important, K increases with particle energy and the equipartition field may be
underestimated significantly.

From a more global point of view the assumption of equipartition seems to be a
very natural one. A thermodynamical system will always distribute the free
energy to all degrees of freedom available, if the system has time to do so. In
cases of accelerated particles diffusing in the large-scale and turbulent
magnetic fields one can expect that at least the turbulent magnetic field (since
it represents three degrees of freedom) is somehow virialized with respect to
any other pressure term, like the cosmic ray pressure. This may not be true for
the ordered magnetic fields, which are supposed to be amplified by the combined
action of differentially rotating shear flows in the disk and some helical
upward and downward motion driven either by cosmic rays pressure or any activity
in the disk. In any case the connection of star formation activity accompanied
by enhanced flux of cosmic rays and the amplification of large-scale magnetic
fields, inherently raises the expectation that magnetic fields should not
exhibit higher pressures than the cosmic rays. Moreover, one would rather expect
to find magnetic fields whose pressure is somewhat lower than the pressure of
the cosmic rays, if the cosmic rays present a source for the galactic dynamo.

The paper is organized as follows: in Sect.~\ref{sect:model} we describe the CR
driven dynamo model and its numerical implementation, in Sect.~\ref{sect:simul}
we present  our simulation setup and describe parameters used in numerical
simulations, in Sect.~\ref{sect:results} we describe results, focusing on the
effect of each parameter on magnetic field amplification rate. We discuss the
final saturated states of models in terms of equipartition between kinetic,
magnetic and CR energies. Finally, in Sect.~\ref{sect:summary} we conclude our
paper.

\section{Description of the model \label{sect:model}}

Similarly as in the papers by Hanasz et al (2004, 2006) we take into account the
following elements of the CR-driven dynamo:

\noindent (1) The cosmic ray component, a relativistic gas, which is described
by the diffusion-advection transport equation (see Hanasz \& Lesch 2003b for the
details of numerical algorithm). The typical values of the diffusion coefficient
found from fitting to CR data (see eg. Strong et al 2007) are $(3 \div 5) \times
10^{28} \cm^2 \s^{-1}$ at energies $\sim 1 \GeV$, and even larger values  $
10^{29} \cm^2 \s^{-1}$ are mentioned (Jokipii 1999), however, we shall use
reduced values in majority of simulations.

\noindent
(2) Following Giaccalone \& Jokipii (1999) and Jokipii (1999) we presume that
cosmic rays diffuse anisotropically along magnetic field lines. The ratio of the
perpendicular to parallel CR diffusion coefficients suggested by these authors
is 5~\%.

\noindent
(3) Localized sources of cosmic rays: supernova remnants exploding randomly in
the disk volume (see Hanasz \& Lesch 2004). We assume that each SN remnant
supplies cosmic rays almost instantaneously, i.e. the comic ray input equal to
10 \% of the canonical SN kinetic energy output ($=10^{51} \erg$) for a single
SN remnant is distributed over several subsequent time-steps.

\noindent
(4) Resistivity of the ISM (see Hanasz et al. 2002, Hanasz \& Lesch 2003a,
Tanuma et al. 2003) responsible for the onset of fast magnetic reconnection and
topological evolution of magnetic field lines. In this paper we apply the
uniform resistivity and neglect the Ohmic heating of gas by resistive
dissipation of magnetic fields.

\noindent
(5) Shearing boundary conditions and tidal forces following the prescription  by
 Hawley, Gammie \&  Balbus (1995) aimed to model differentially rotating disks
in the local approximation.

\noindent
(6) Realistic vertical disk gravity following the model of ISM in the Milky Way
by Ferriere (1998).

The set of equations describing the model of the CR-driven dynamo includes
resistive MHD and the cosmic ray transport equations (see Hanasz et al., 2004):

\begin{equation}\pder{\rho}{t}{}+ \bm{\nabla} \cdot (\rho \bm{V}) = 0, \label{eqofconti}
\end{equation}
\begin{equation}
\pder{e}{t}{} +\bm{\nabla}\cdot \left( e \bm{V}\right)
    = - p \left( \bm{\nabla} \cdot \bm{V}
\right), \label{enereq}
\end{equation}
\begin{eqnarray}
\pder{\bm{V}}{t}{} + (\bm{V} \cdot \bm{\nabla})\bm{V}
   = -\frac{1}{\rho} \bm{\nabla}
   \left(p + p_{\rm cr} + \frac{B^2}{8\pi}\right)   \nonumber \\
   + \frac{\bm{B \cdot \nabla B}}{4 \pi \rho}
      -2 \bm{\Omega} \times \bm{v}
   + 2 q {\Omega}^2 \rm x \hat{\bm{e}}_x + g_z(z)\, \hat{\bm{e}}_z,
   \label{eqofmot}
\end{eqnarray}

\begin{equation}
\pder{\bm{B}}{t}{} = \bm{\nabla} \times \left( \bm{V} \times \bm{B}\right)
+ \eta \Delta \bm{B}
\label{indeq},
\end{equation}

\begin{equation}
p=(\gamma-1) e, \quad \gamma=5/3
\end{equation}
where $\rm q=- \rm{d~ln\Omega/d~lnR}$ is the shearing parameter, $\rm R$ is the
distance to galactic center, $\eta $ is the resistivity, $\gamma$ is the
adiabatic index of thermal gas, the gradient of cosmic ray pressure  $\nabla
p_{cr}$ is included in the equation of motion (see eg. Berezinski et al. 1990)
and other symbols have their usual meaning. The uniform resistivity is included
only in the induction equation (see Hanasz et al. 2002). The thermal gas
component is currently treated as an adiabatic medium.

The transport of the cosmic ray component is described by the
diffusion-advection equation (see eg. Berezinski et al. 1990, Schlickeiser \&
Lerche 1985)
\begin{equation}
\pder{e_{\rm cr}}{t}{} +\bm{\nabla }\left( e_{\rm cr} \bm{V}\right)
= \bm{\nabla} \left(\hat{K} \bm{\nabla} e_{cr} \right)
- p_{\rm cr} \left( \bm{\nabla} \cdot \bm{V} \right)
+ Q_{\rm SN}, \label{diff-adv-eq}
\end{equation}
where $Q_{\rm SN}$ represents the source term for the cosmic ray energy density:
the rate of production of cosmic rays injected locally in the SN remnants and
\begin{equation}
p_{\rm cr}=(\gamma_{\rm cr}-1) e_{\rm cr}, \quad \gamma_{\rm cr}=14/9.
\end{equation}
The adiabatic index of the cosmic ray gas $\gamma_{\rm cr}$ and the formula for
diffusion tensor
\begin{equation}
K_{ij} = K_{\rm \perp} \delta_{ij} + (K_\parallel - K_{\rm \perp}) n_i n_j,
\quad n_i = B_i/B,
\label{diftens}
\end{equation}
are adopted following the argumentation by Ryu et al. (2003).

\section{Numerical simulations \label{sect:simul}}

\subsection{Simulation setup}

In this paper we present a series of recent numerical simulations, whose aim is
to search for the most favorable conditions for magnetic field amplification by
means of the CR-driven dynamo.  The presented numerical simulations were
performed with the aid of Zeus-3D MHD code (Stone and Norman 1992 a,b) extended
with additions to the standard algorithm, that correspond to items (1) - (6) of
Sect.~\ref{sect:model}, i.e.  the cosmic ray component, treated as a fluid and
described by the diffusion-advection equation, including anisotropic CR
diffusion tensor and cosmic ray sources  -- supernova remnants  exploding
randomly in the disk  volume,  resistivity of the ISM leading to magnetic
reconnection, shearing-periodic boundary conditions, rotational pseudo-forces
and a realistic vertical disk gravity.

All simulations are performed in a Cartesian domain of size 0.5 kpc $\times$ 1
kpc $\times$ 2 kpc in $x$, $y$, $z$ coordinates, corresponding to radial,
azimuthal and vertical directions, respectively. The basic resolution of the
numerical grid is 50 $\times$ 100 $\times$ 400 grid cells in $x$, $y$ and $z$
directions, respectively, and for a smaller sample of simulations performed with
larger values of CR diffusion coefficients the grid resolution is 25\ $\times$
50 $\times$ 200 grid cells. The boundary conditions are sheared-periodic in
coordinate $x$, periodic in coordinate $y$ and outflow on outer $z$-boundaries,
with $e_{cr} = 0$ at the domain boundaries. The positions of SN are chosen
randomly, with a uniform distribution in $xy$ coordinates and Gaussian
distribution in $z$ coordinate.

The initial density distribution results from integration of the
hydrostatic equilibrium equation, for the vertical gravity model of Ferriere
(1998) and with the assumption of constant gas temperature across the disk,
equal approximately 6000 K corresponding to the sound speed equal to about 7
km/s. The integration procedure finds hydrostatic equilibrium for a given
gas column density treated as an input parameter.

The magnetic field strength, incorporated in the initial hydrostatic
equilibrium of gas,  is defined through the parameter $\alpha$ denoting the
ratio of initial magnetic to gas pressures. The initial cosmic ray pressure is
equal to the initial gas pressure.

The CR energy supplied to the system in SN remnants,  randomly distributed 
around the disk midplane, implies that the CR pressure gradient force
accelerates a vertical wind of thermal gas. To prevent significant mass
losses from the computational domain, due to the vertical wind, we compensate
the mass-loss $\Delta m$ after each timestep. The compensation mass
is supplied as a mass source term, which is proportional to
the initial mass distribution

\begin{equation}
\Delta \rho (x,y,z) = \frac{\Delta m}{m_{\rm tot}} \rho_0 (x,y,z),
\label{eq:mass-compens}
\end{equation}
where $m_{\rm tot}$ is the total gas mass in the computational domain and 
$\rho_0 (x,y,z)$ is the initial density distribution, resulting from  
the integration procedure of the hydrostatic equilibrium equation.

\subsection{Simulation parameters}

\begin{table*}
\scriptsize
 \begin{center}
  \begin{tabular}{|c|c|c|c|c|c|c|c|c|}
   \hline\hline
   Simulation & $\alpha$ & $\eta$ &  $\Rm$ &   $f_{SN}$       & $f_{SN}$ modul. & $K_\paral$    & $K_\perp$  & $n_x\times n_y\times n_z$  \\
          &	  &$[\pc^2\Myr^{-1}]$ & &$[\kpc^{-2}\Myr^{-1}]$ &  (Y/N)        & $[\pc^2\Myr^{-1}]$ & $[\pc^2\Myr^{-1}]$ & \\
   \hline\hline
    A1    & $10^{-4}$&	   0 &	 $\infty$     &  130&	   N	    &	$1\times10^4$	&   $1\times10^3$ &$100\times50\times400$     \\
    A2    & $10^{-4}$&	   1 & $5\times 10^4$ &130&	   N	    &	$1\times10^4$	&   $1\times10^3$ &$100\times50\times400$     \\
    A3    & $10^{-4}$&    10 & $5\times 10^3$ &		130&	   N	    &	$1\times10^4$	&   $1\times10^3$ &$100\times50\times400$     \\
    A4    & $10^{-4}$&   100 & $5\times 10^2$ &		130&	   N	    &	$1\times10^4$	&   $1\times10^3$ &$100\times50\times400$     \\
    A5    & $10^{-4}$&  1000 & $5\times 10^1$ &		130&	   N	    &	$1\times10^4$	&   $1\times10^3$ &$100\times50\times400$     \\
   \hline
    B1    & $10^{-4}$&	   0 & $\infty$  &		130&	   Y	    &	$1\times10^4$	&   $1\times10^3$ &$100\times50\times400$     \\
    B2    & $10^{-4}$&	   1 & $5\times 10^4$ &		130&	   Y	    &	$1\times10^4$	&   $1\times10^3$ &$100\times50\times400$     \\
    B3    & $10^{-4}$&    10 & $5\times 10^3$ &		130&	   Y	    &	$1\times10^4$	&   $1\times10^3$ &$100\times50\times400$     \\
    B4    & $10^{-4}$&   100 & $5\times 10^2$ &		130&	   Y	    &	$1\times10^4$	&   $1\times10^3$ &$100\times50\times400$     \\
    B5    & $10^{-4}$&  1000 & $5\times 10^1$ &		130&	   Y	    &	$1\times10^4$	&   $1\times10^3$ &$100\times50\times400$     \\
   \hline
    C1    & $10^{-4}$&   100 & $5\times 10^2$ &	15 &	   Y	    &	$1\times10^4$	&   $1\times10^3$ &$100\times50\times400$     \\
    C2    & $10^{-4}$&   100 & $5\times 10^2$ &		30 &	   Y	    &	$1\times10^4$	&   $1\times10^3$ &$100\times50\times400$     \\
    C3    & $10^{-4}$&   100 & $5\times 10^2$ &		60 &	   Y	    &	$1\times10^4$	&   $1\times10^3$ &$100\times50\times400$     \\
    C4    & $10^{-4}$&   100 & $5\times 10^2$ &		250&	   Y	    &	$1\times10^4$	&   $1\times10^3$ &$100\times50\times400$     \\
    C5    & $10^{-4}$&   100 & $5\times 10^2$ &		500&	   Y	    &	$1\times10^4$	&   $1\times10^3$ &$100\times50\times400$     \\
  \hline
    D1    & $10^{-4}$&   100 & $5\times 10^2$ &		130&	   N	    &	$1\times10^4$	&   $1\times10^3$ &$50\times25\times200$     \\
    D2    & $10^{-4}$&   100 & $5\times 10^2$ &		130&	   Y	    &	$1\times10^4$	&   $1\times10^3$ &$50\times25\times200$     \\
   \hline
    E1    & $10^{-2}$&   100 & $5\times 10^2$ &		130&	   N	    &	$1\times10^4$	&   $3\times10^3$ &$50\times25\times200$     \\
    E2    & $10^{-2}$&   100 & $5\times 10^2$ &		130&	   N	    &	$3\times10^4$	&   $1\times10^3$ &$50\times25\times200$     \\
    E3    & $10^{-2}$&   100 & $5\times 10^2$ &		130&	   N	    &	$3\times10^4$	&   $3\times10^3$ &$50\times25\times200$     \\
    E4    & $10^{-2}$&   100 & $5\times 10^2$ &		130&	   N	    &	$3\times10^4$	&   $1\times10^4$ &$50\times25\times200$     \\
    E5    & $10^{-2}$&   100 & $5\times 10^2$ &		130&	   N	    &	$1\times10^5$	&   $1\times10^3$ &$50\times25\times200$     \\
    E6    & $10^{-2}$&   100 & $5\times 10^2$ &		130&	   N	    &	$1\times10^5$	&   $3\times10^3$ &$50\times25\times200$     \\
    E7    & $10^{-2}$&   100 & $5\times 10^2$ &		130&	   N	    &	$1\times10^5$	&   $1\times10^4$ &$50\times25\times200$     \\
   \hline
  \end{tabular}
 \end{center}
 \caption{Parameters of simulations presented in this paper. Subsequent columns
show: simulation name, initial ratio of magnetic to gas pressure $\alpha$,
magnetic diffusivity $\eta$, magnetic Reynolds number $\Rm$, surface frequency of SN explosions, presence of
SN-rate modulation, parallel $K_\paral$ and perpendicular $K_\perp$ CR diffusion
coefficients,  and  grid resolution in $x$, $y$ and $z$ directions. }

 \label{table-params}
\end{table*}

The basic input parameters, resulting from the assumed model are: the vertical
gravity profile, local value of the galactic rotation and shear,  gas column
density and supernova rate.  We adopt these parameters from the global model of
ISM in Milky Way (Ferriere 1998) for the galactocentric radius $R=5 \kpc$, where
angular velocity is $\Omega = 0.05 \Myr^{-1} $, gas column density  $\Sigma= 27
\times 10^{20} \cm^{-2}$ and the realistic vertical gravity given by formula
(36) in the aforementioned paper.  The values of gas column density correspond
in our simulations to the total density of all gas components in Ferriere
(1998), while SN-rate is the rate of type II supernovae. We assume for
simplicity that all SN explosions appear as single supernovae, and that vertical
distribution of SN explosions is Gaussian, with a fixed half-width equal to 100
pc.

In addition to the mentioned relatively well established local disk parameters,
there is a group of less known quantities like effective magnetic diffusivity,
CR diffusion coefficients and efficiency of conversion of SN kinetic energy into
cosmic ray energy.  We assume the standard 10\% value of kinetic CR energy
conversion efficiency and the magnetic diffusivity and CR diffusion coefficients
varying in a wide range. 

In this paper we present the results of five simulation series A--E. The summary
of all variable simulations parameters for the whole set of simulations is
presented in Table~\ref{table-params}.

In simulation series A (runs A1 - A5) we examine effects of  magnetic
diffusivity variations on magnetic field amplification by applying  $\eta$ in
the range $0 \div 10^3 \pc^2 \Myr^{-1}$ corresponding to $0 \div 3\times 10^{26}
\cm^2 \s^{-1}$ in CGS units. We define the magnetic Reynolds number  $\Rm =
L_y^2 \Omega/\eta$ for reference, as in Gressel et al. (2008b), where $L_y=1000
\pc$ is the domain size in $y$ direction. Moreover we assume continuous and
time-invariable supply of CRs in SN remnants. The simulation runs A1, A2, A3 are
the same as the runs B, C, and D, respectively, discussed by Otmianowska-Mazur
et al (2007).  We note, for comparison,  that the commonly adopted value of
turbulent diffusivity in the ISM is $\eta_{\rm turb}\simeq 1/3 v_{\rm turb}
L_{\rm turb} \sim 10^{26} \cm^2\s^{-1} $ ($\simeq 1/3 \times 10^3
\pc^2\Myr^{-1}$) for $v_{\rm turb} = 10 \km \s^{-1}$ and $ L_{\rm turb} = 100
\pc$.  We note that the adopted values of magnetic diffusivity exceed the value
corresponding to the Spitzer resistivity ($\simeq 10^8 \cm^2 \s^{-1}$, see
Parker 1992) by 15--18 orders of magnitude. The relative smallness of the
Spitzer resistivity implies  that an anomalous resistivity, considered as a
subscale phenomenon, has to be invoked in order to explain dissipation of the
small scale magnetic fluctuations in the ISM. Following Parker (1992), we assume
that reconnection rates in the ISM are comparable to those predicted by the
Petscheck's fast reconnection model, i.e. the magnetic cutting speeds are of the
order of $v_A/\log({\rm R_M})$ rather than $v_A/\sqrt{\rm R_M}$ typical for the
slow Parker-Sweet reconnection model, where ${\rm R_M}$ is the Lundquist number
or the magnetic Reynolds number. 

In the simulation series B (runs B1 - B5) we apply the same range of magnetic
diffusivity values, but CRs supply is modulated in a manner mimicking passages
of subsequent spiral arms, regulating the star formation rate and subsequently
SN-rate. The effect of spiral arms is modelled (see Hanasz et al. 2006) by
supplying cosmic rays in intermittent periods of 25 Myr, with SN rate equal to
$4 \times$ the  reference  $f_{\rm SN}$, and followed by periods of 75 Myr
without any SN activity.  The time-averaged supernova rate is in this case equal
to the reference  $f_{\rm SN}$.

In the simulation series C (runs C1-C5) we apply a constant magnetic diffusivity
$\eta = 100 \pc^2 \Myr^{-1}$ and vary the surface frequency of SN explosions in
the range of $15-500 \, \kpc^{-2} \Myr^{-1}$, assuming modulated CR supply as in
the simulation series B. In the simulation series D (runs D1 - D2) we repeat
simulations A4 and B4, respectively,  with the grid resolution reduced twice in
each spatial direction.

Due to the CFL timestep limitation of the currently used explicit algorithm of
CR diffusion, the applied values of CR diffusion coefficients are scaled down
with respect to the realistic values. The timestep limitation ensuring stability
of explicit numerical schemes applied to the diffusion equation is $\Delta t
\leq 0.5 (\Delta x)^2/K$, where $K$ is the diffusion coefficient. The timestep
becomes prohibitively short when the diffusion coefficient is very large or the
spatial step is too small. For this reason the CR diffusion coefficient has been
reduced in simulation series A-D, by about one order of magnitudes, with respect
to the mentioned realistic values $3\div 6 \times 10^{28} \cm^2\s^{-1} \simeq
1\div 2 \cdot 10^5 \pc^2 \Myr^{-1}$. The fiducial values of the parallel and
perpendicular diffusion coefficients applied in simulation series A--D  are
respectively:  $K_\paral = 1\times10^4 \pc^2\Myr^{-1}\simeq 3\times 10^{27}
\cm^2 \s^{-1}$ and  $K_\perp = 1\times10^3\simeq 3\times 10^{26} \cm^2 \s^{-1}$.

Finally, in the simulation series E (runs E1 - E6) we increase the parallel
$K_\paral$ and perpendicular $K_\perp$ CR diffusion coefficients by factors 3
and 10, with respect to the fiducial values, in order to examine magnetic field
amplification for more realistic magnitudes of these quantities. In this way we
apply realistic CR diffusion coefficients in a few single simulation runs, yet
the maximum CR diffusion coefficients used do not reach the upper range of
realistic values, of the order of $10^{29} \cm^2 \s^{-1}$, mentioned in the literature.

\section{Results \label{sect:results}}

\begin{figure*}
\centerline{
  \includegraphics[height=0.45\textheight]{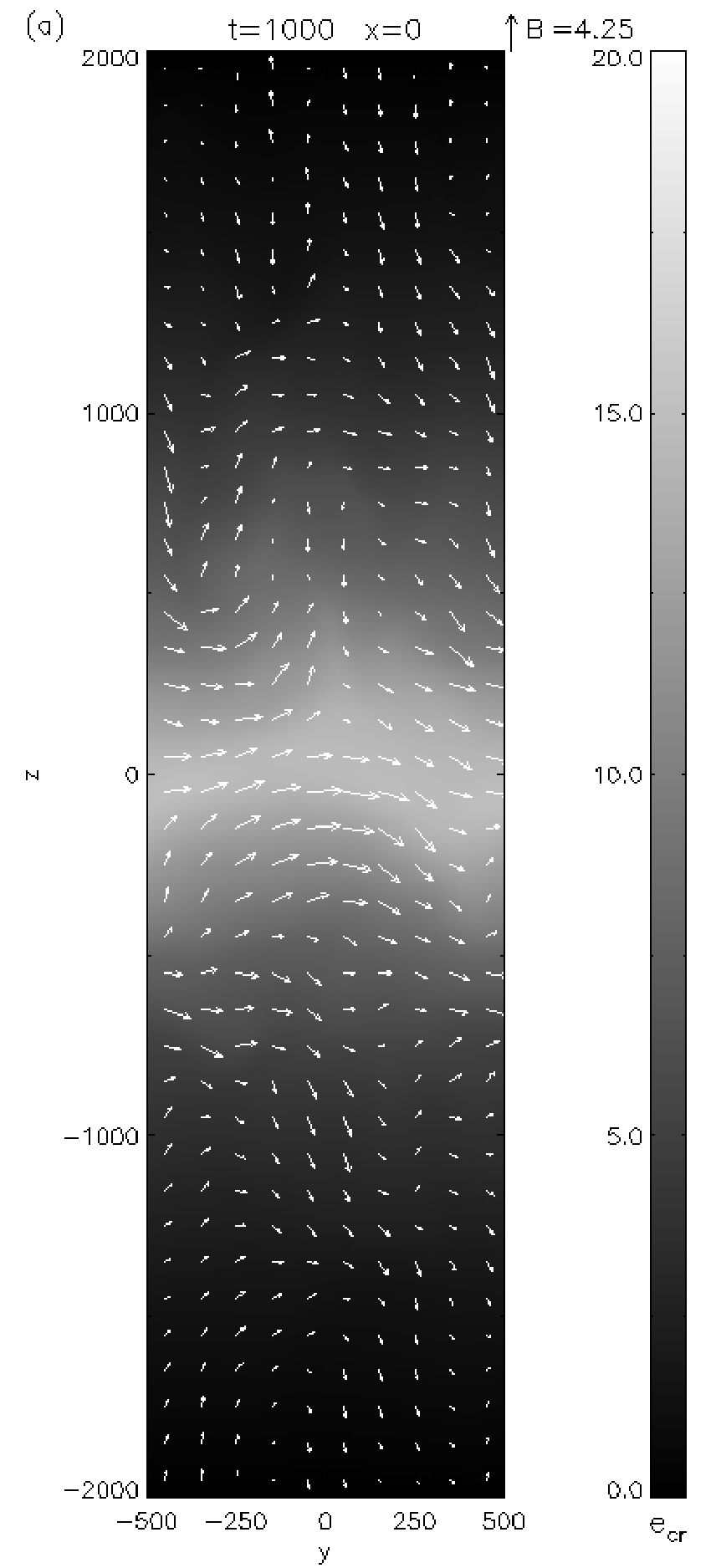}
  \includegraphics[height=0.45\textheight]{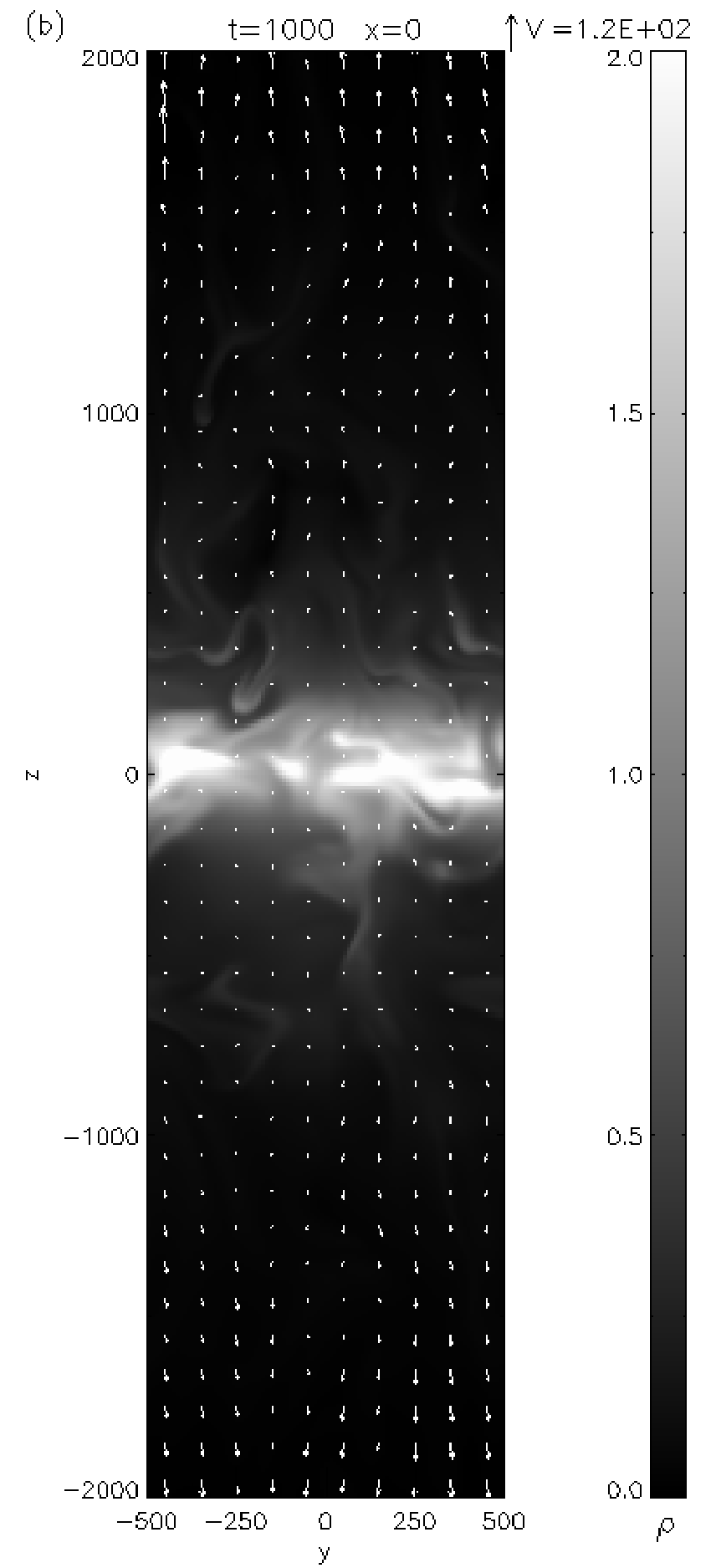}
  \includegraphics[height=0.45\textheight]{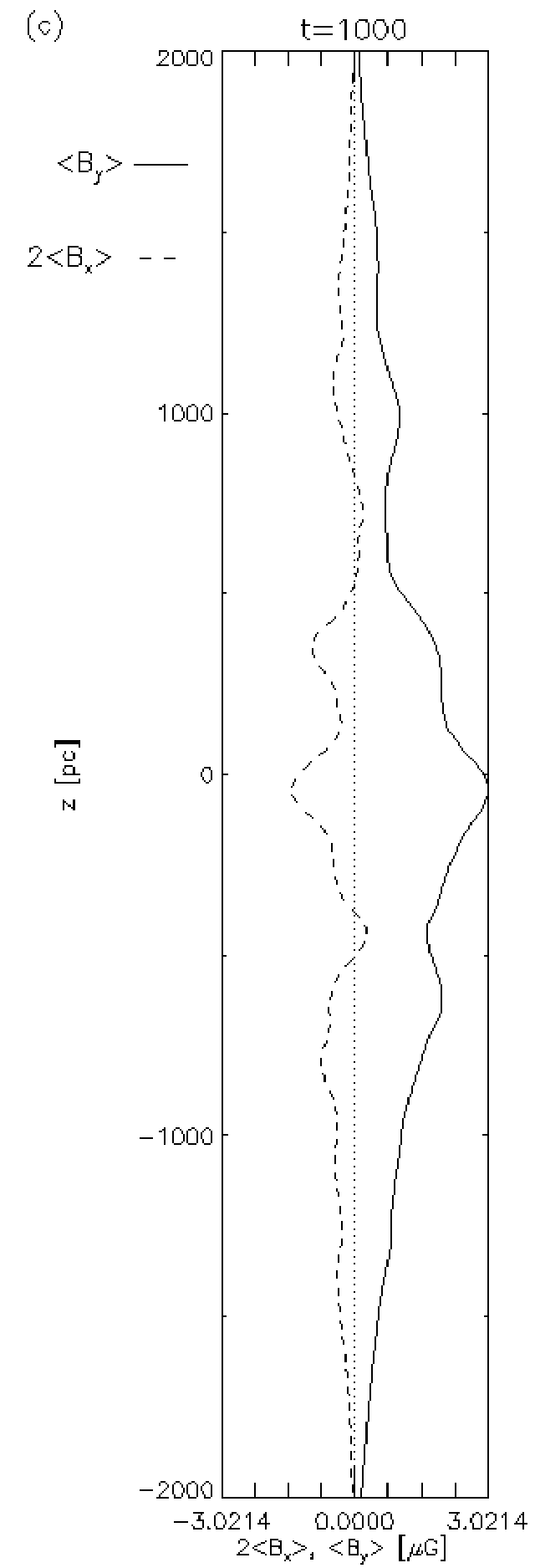}
  \includegraphics[height=0.45\textheight]{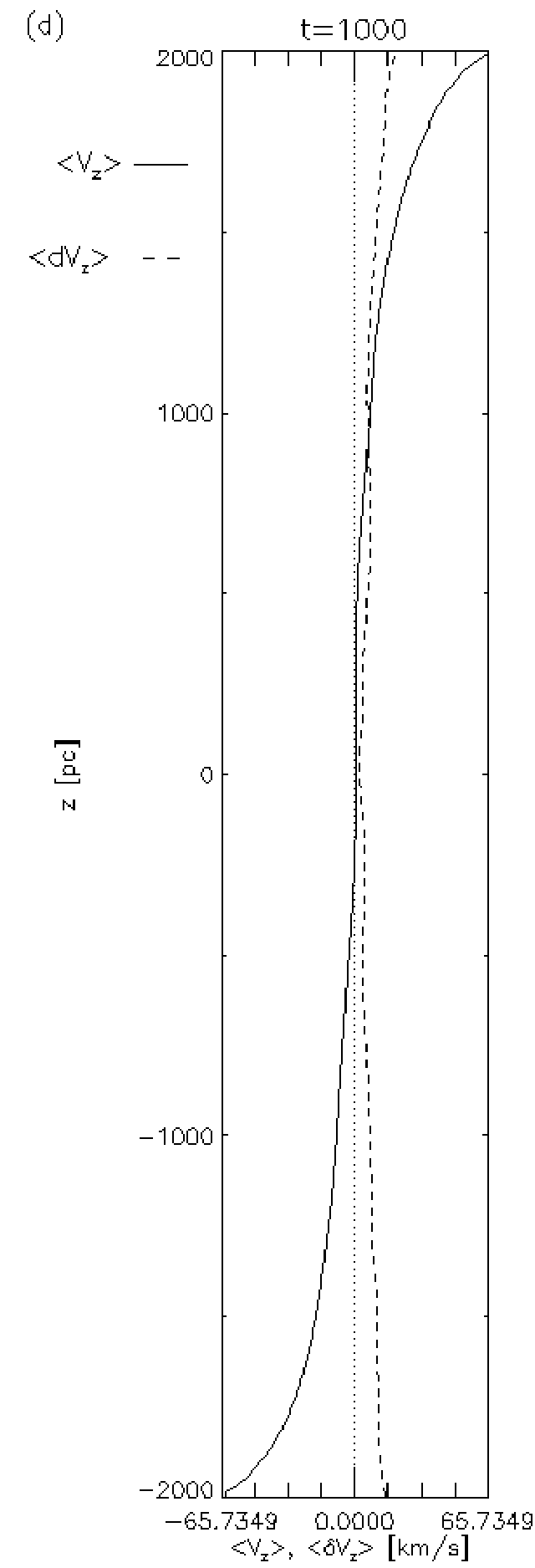}
}
\caption{Exemplary plots illustrating the state of the system at $t=1000 \Myr$
for simulation A4. In the first two panels we present slices through the
computational volume in the $yz$-plane for $x =0$. Panel (a) shows cosmic ray
energy density with vectors of magnetic field, panel (b) shows gas density with
velocity vectors. In panel (c) we plot horizontally averaged $x$ and $y$
components of magnetic field, and in panel (d) horizontally averaged vertical
velocity component and its fluctuations.}
\label{fig:slices}
\end{figure*}

In Fig.~\ref{fig:slices} we show the distribution of cosmic ray gas together
with magnetic field vectors (panel (a)), and thermal gas density together with
gas velocity vectors (panel (b)) in the $yz$-slice taken for $x=0$  at $t=1000$
Myr.

One can notice in panel (a) that the dominating horizontal magnetic field
component in the disk volume is undulated in a manner resembling the effects of
Parker instability. The cosmic ray energy density is well smoothed by the
diffusive transport in the computational volume. The vertical gradient of the
cosmic ray energy density is maintained by the supply of cosmic rays around the
equatorial plane in the disk in the presence of vertical gravity. The cosmic ray
energy density is expressed in units in which the thermal gas energy density
corresponding to $\rho=1 \cm^{-3}$ and the isothermal sound speed $c_{s}= 7
\km\s^{-1}$ is equal to 1.  The velocity field together with gas density is
shown in panel (b).  It is apparent that the distribution of gas is
significantly less smooth that the distribution of cosmic rays.  

In order to examine the structure of the large-scale field we show in panel (c)
the horizontally averaged magnetic field components $\langle B_x (z)\rangle$ and and $\langle
B_y (z)\rangle$.  A striking property of the mean magnetic
field configuration is the almost exact coincidence of peaks of the oppositely
directed radial and  azimuthal field components. This feature resembles to the
standard picture of an $\alpha\omega$-dynamo: the azimuthal mean magnetic
component is generated from the radial one and vice versa (see Lesch and Hanasz
(2003) for a corresponding simple analytical model).

In panel (d) we show the horizontally averaged vertical velocity component
$\langle V_z \rangle$  and its fluctuations $\langle \delta V_z \rangle$. It is
apparent that the bulk speeds of the wind, driven by the vertical gradient of CR
pressure,  reach $65 \pm 15 \km/\s$ at $z=\pm 2\kpc$. Vertical systematic winds
blowing with bulk speeds, comparable to rotational galactic velocities,
influence large-scale structures of galactic magnetic fields and are observed in
external starburst galaxies like NGC 253 (see Beck 2007, Hessen et al. 2008).

By varying the parameters discussed in the previous section we intend to
determine the regions of parameter space in which  the magnetic field
amplification is the most efficient. The amplification of the regular magnetic
field is identified with the amplification of the total magnetic energy in the
computational domain, associated with  the amplification of the azimuthal
magnetic flux. The magnetic flux shown in subsequent plots represents an
azimuthal flux averaged over all $xz$ slices (spread in  $y$ direction) through
the discretized computational domain.   In the subsequent subsections we present
the parameter study of the CR-driven dynamo, focusing on the efficiency of
magnetic field amplification and the issue of equipartition between magnetic,
kinetic, and CR energies.

\subsection{Dependence of magnetic field amplification on magnetic diffusivity
\label{sect:eta}}

\begin{figure*}
\centerline{\qquad\includegraphics[width=0.45\textwidth]{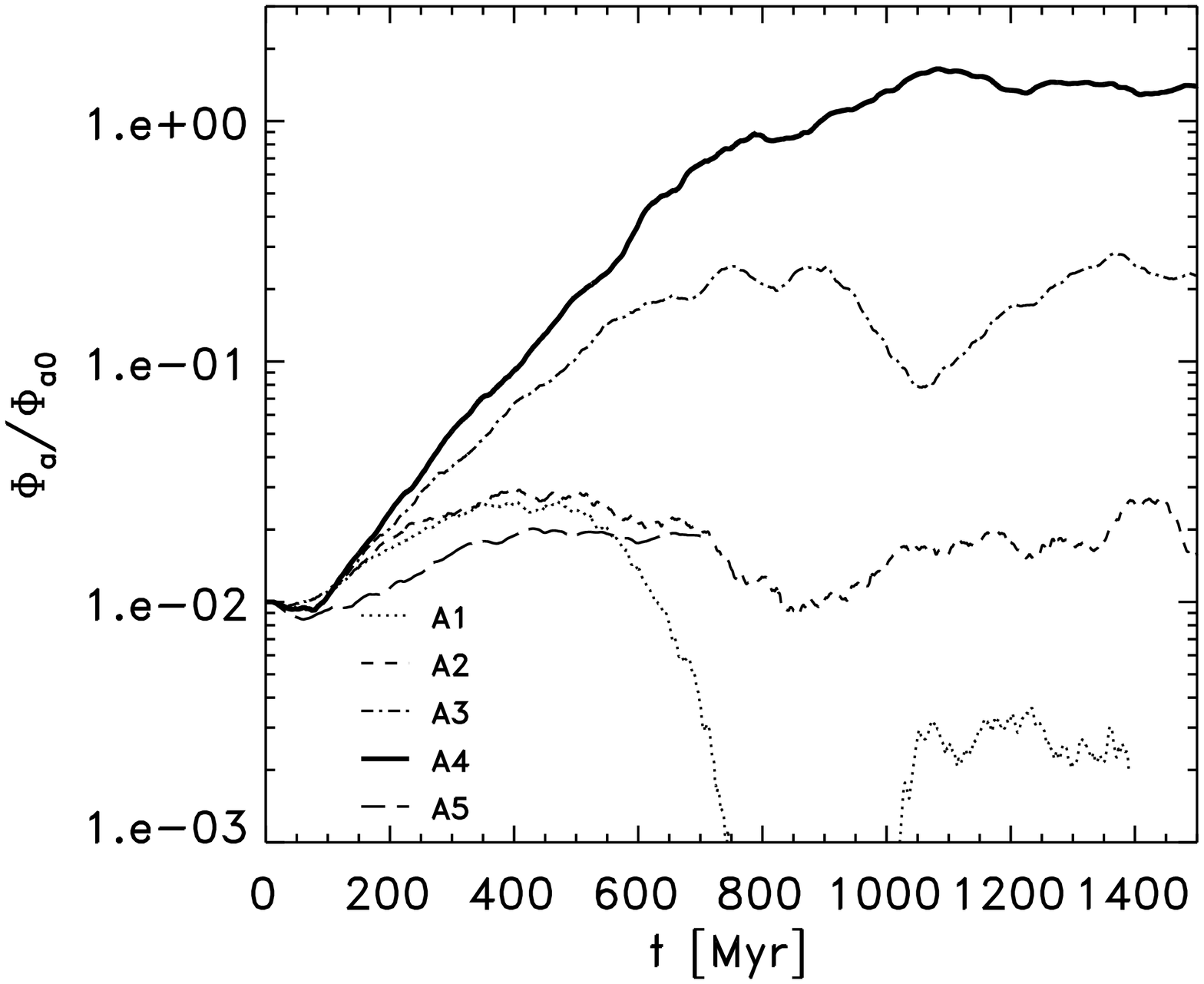}
            \qquad\includegraphics[width=0.45\textwidth]{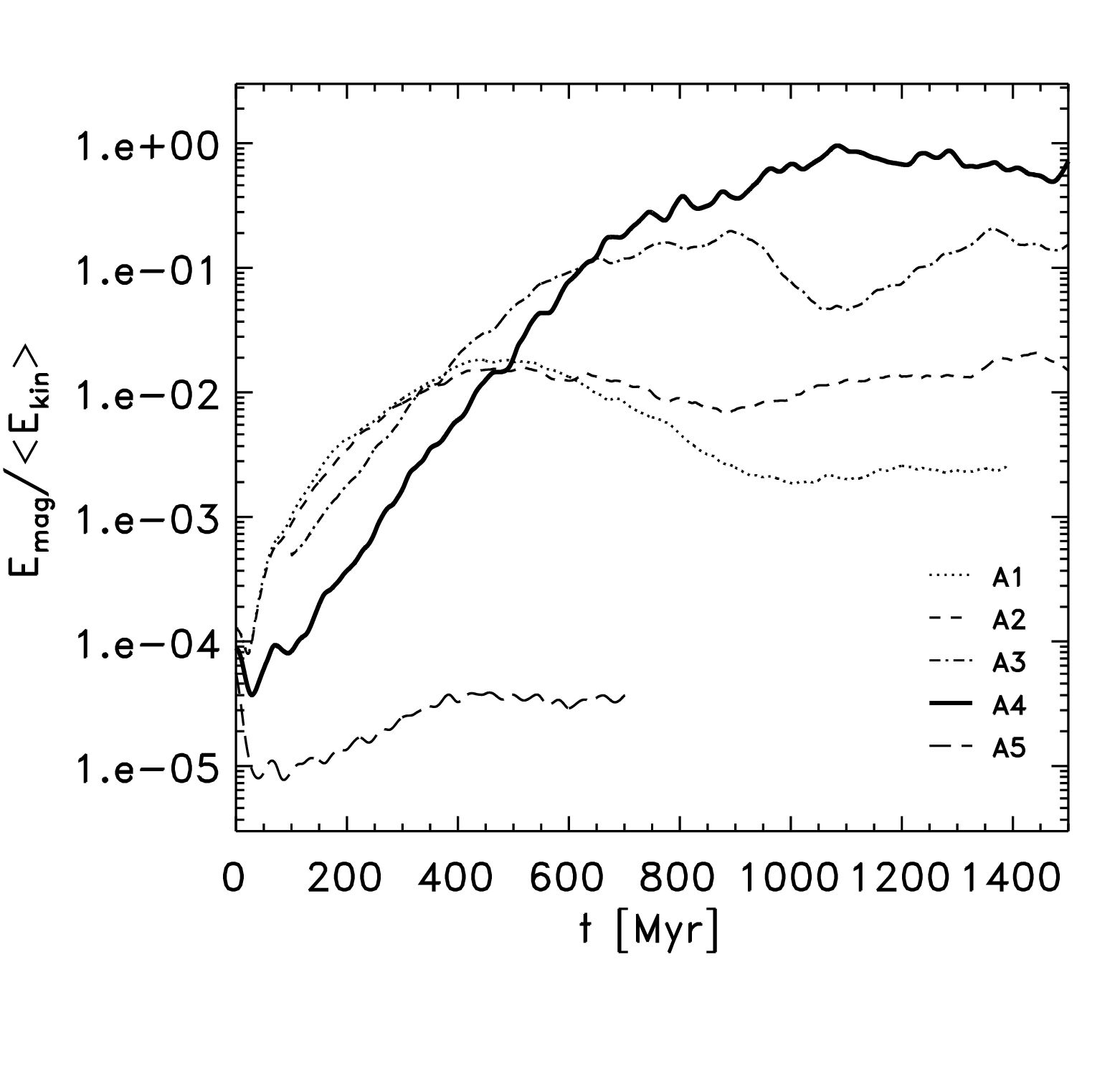}}
\caption{Time evolution of azimuthal  magnetic flux and total magnetic energy
for different values of magnetic diffusivity in simulation series A. The curves represent
respectively cases of $\eta=0$ (A1), $\eta=1$ (A2), $\eta=10$ (A3), $\eta=100$ (A4) 
and $\eta=1000$ (A5) in units $\pc^2 \Myr^{-1}$.}
\label{fig:eta}
\end{figure*}

As a first step in our parameter study of the cosmic-ray driven dynamo we
examine, in simulation series A, the effect of magnetic diffusivity on the
efficiency of magnetic field amplification.  Time evolution of magnetic energy
and magnetic flux are shown in Fig.~\ref{fig:eta}. Magnetic flux plotted in the
left panel of Fig.~\ref{fig:eta} is scaled in the following way.  The initial
magnetic field induction is defined by the parameter $\alpha = p_{mag}/p_{gas}$,
shown in the second column of Table.~\ref{table-params},
where we apply Parker's convention to assign the inverse of plasma beta as $\alpha$. The adopted values of
$\alpha$ are $10^{-4}$ and $10^{-2}$ in different simulations, while $\alpha =
1$ means magnetic pressure equal to the thermal gas pressure. We scale magnetic flux in
such a way that $\alpha = 1$ corresponds to the azimuthal magnetic flux $\Phi_a
= 1$. The total magnetic energy plotted in the right panel of Fig.~\ref{fig:eta}
is scaled with respect to the time-averaged total kinetic energy in the
computational domain. The latter quantity appears to fluctuate around a mean
value, which is practically time-invariant for all simulation runs, thus we find
this kind of scaling convenient. The scaling described above will be applied to
all subsequent plots of magnetic flux and magnetic energy.

The evolution of magnetic energy and magnetic flux in models of simulation series
A, represented by different curves in Fig.~\ref{fig:eta}, demonstrate that magnetic
field amplification strongly depends on magnetic diffusivity. In the case of
vanishing explicit resistivity, magnetic energy grows up by about 2.5 orders of
magnitude during the first 500 Myr, but magnetic flux is amplified only by a factor
of 3 during this period and it fades later on. This effect can be interpreted as a
predominant growth of the small-scale turbulent magnetic field component with a
little contribution of large-scale magnetic field amplification (see
Otmianowska-Mazur et al. 2007  for a more extended analysis of the simulations
presented in this section). One should remember, however, that numerical
resistivity, always present in numerical MHD simulations, may influence to some
extent the behavior of the simulation run A, corresponding to $\eta=0$. The amount
of numerical magnetic diffusivity  $ \eta \simeq 0.7 \pc^2 \Myr^{-1}$ has been
quantified, for the present grid resolution, on the base of Parker instability
simulations by Kowal et al (2003).

When magnetic diffusivity is increased up to $\eta=100 \ \pc^2 \Myr^{-1}$ the
efficiency of magnetic field amplification increases.   For $\eta=100 \pc^2
\Myr^{-1}$ (run A4) the growth of magnetic flux persists until $t= 1000\, \Myr$
and saturates thereafter. For smaller values of $\eta$ the growth rate is
smaller and the maximum values of magnetic flux are smaller than those attained
for $\eta=100 \pc^2 \Myr^{-1}$.  In cases of vanishing or small explicit
diffusivity (runs A1 and A2), magnetic energy grows up initially faster than in
the case of larger resistivity (runs A3 and A4).  This behavior means that low
resistivity enables initially faster growth of the random magnetic field
component, while for larger resistivity random magnetic fields are quickly
dissipated. The growth of the total magnetic energy follows closely, in the
latter case, the growth of the mean magnetic flux. It is also apparent that
amplification of magnetic flux and magnetic energy for $\eta=1000 \pc^2
\Myr^{-1}$ (run A5) is significantly reduced with respect to the other runs.

\begin{figure}
\centerline{\qquad\includegraphics[width=0.45\textwidth]{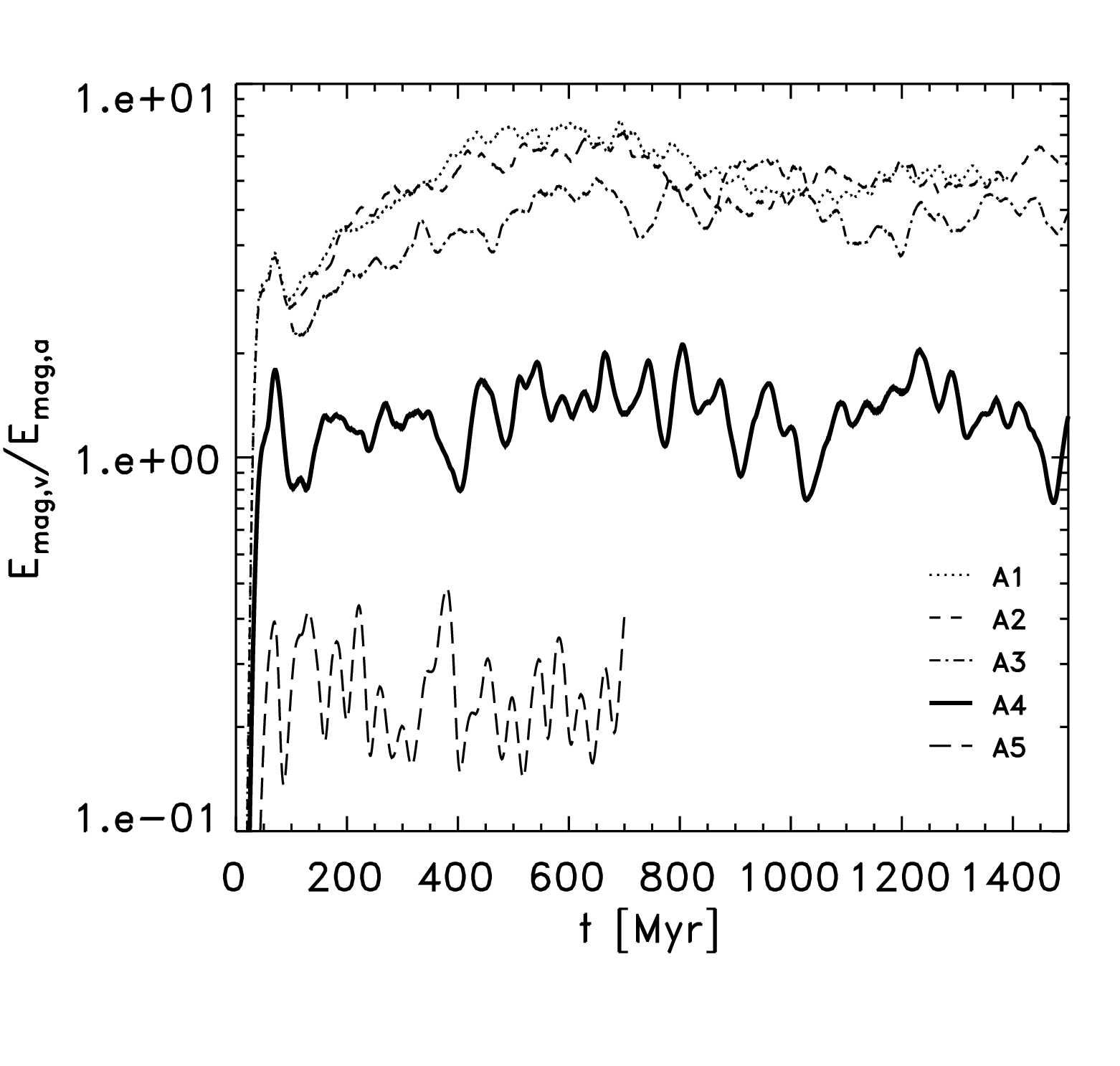}}
\caption{Time evolution of the ratio of energies of vertical to horizontal
magnetic field components for different values of magnetic diffusivity in the
simulation series A. Line assignments are the same as in Fig.~\ref{fig:eta}}
\label{bv-bz-eta}
\end{figure}

In order to explain the physical mechanism that controls magnetic field
amplification through the magnitude of magnetic diffusivity we plot the ratio of
total (volume integrated) energies of vertical to azimuthal magnetic field
components in Fig.~\ref{bv-bz-eta}. It is apparent that the energy of vertical
magnetic field dominates in all lower magnetic diffusivity runs A1-A3 ($\eta =
0, 1$ and $10 \pc^2\Myr^{-1}$) and is comparable to the energy of azimuthal
magnetic field in run A4 ($\eta = 100 \pc^2\Myr^{-1}$) providing the strongest
magnetic field amplification. Among all simulations of series A only simulation
A4 reaches energetic equipartition between magnetic field and gas kinetic energy
as a result of the amplification process.

On the other hand, in the case of large magnetic diffusivity (run A5, $\eta=1000
\pc^2\Myr^{-1}$) energy of the vertical magnetic field component remains much
smaller than energy of the azimuthal magnetic field and magnetic field
amplification does not occur. This fact can be interpreted by resistive damping
of the undulatory mode of the Parker instability in favor of the interchange
mode, which does not contribute to the dynamo action.  The above finding
indicates that {\em the most favorable conditions for magnetic field
amplification corresponds to approximately equal energies in vertical and
azimuthal magnetic fields in the case of buoyancy driven dynamo.}

\begin{figure}
\centerline{\includegraphics[width=\columnwidth]{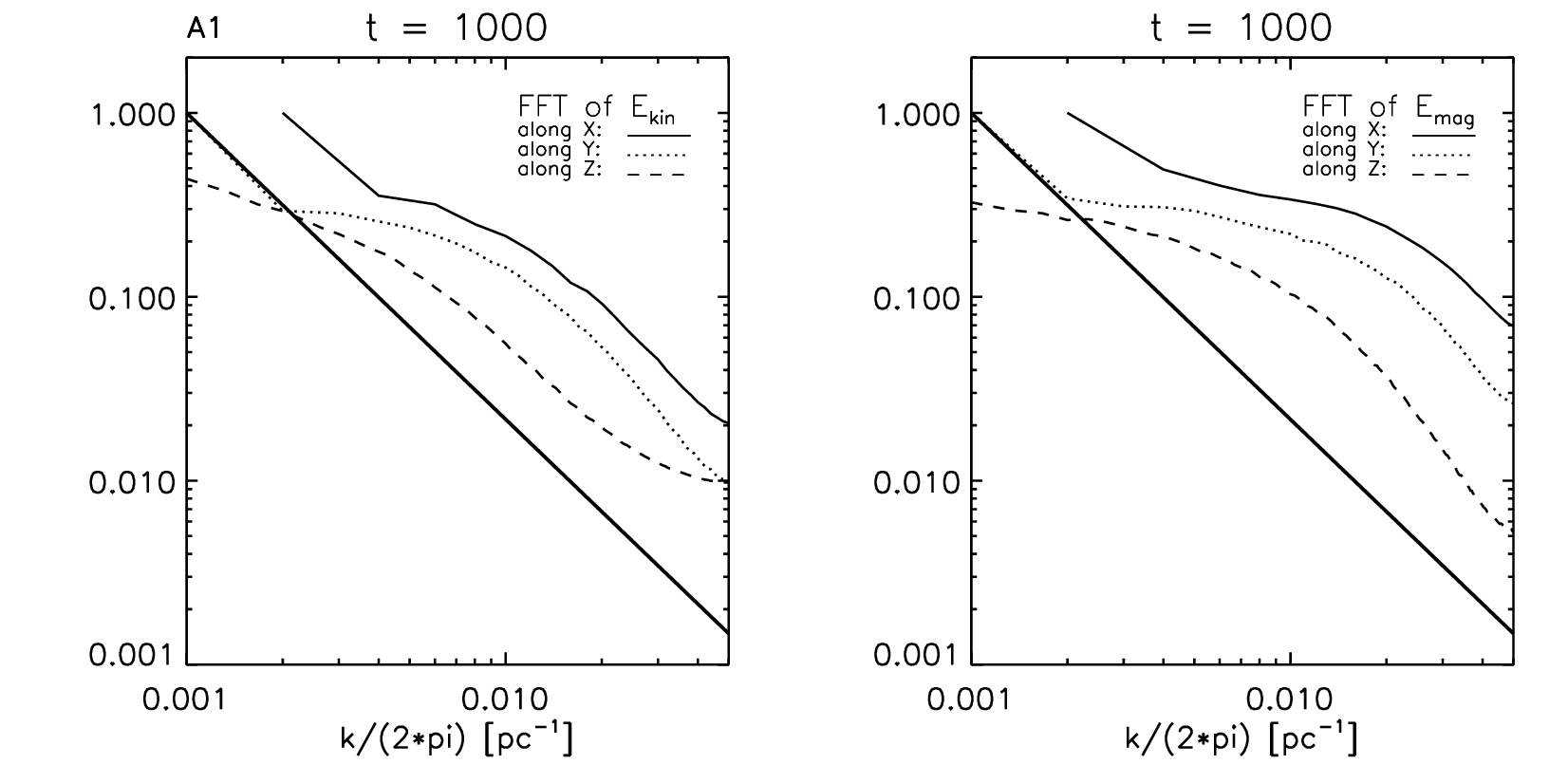}}
\centerline{\includegraphics[width=\columnwidth]{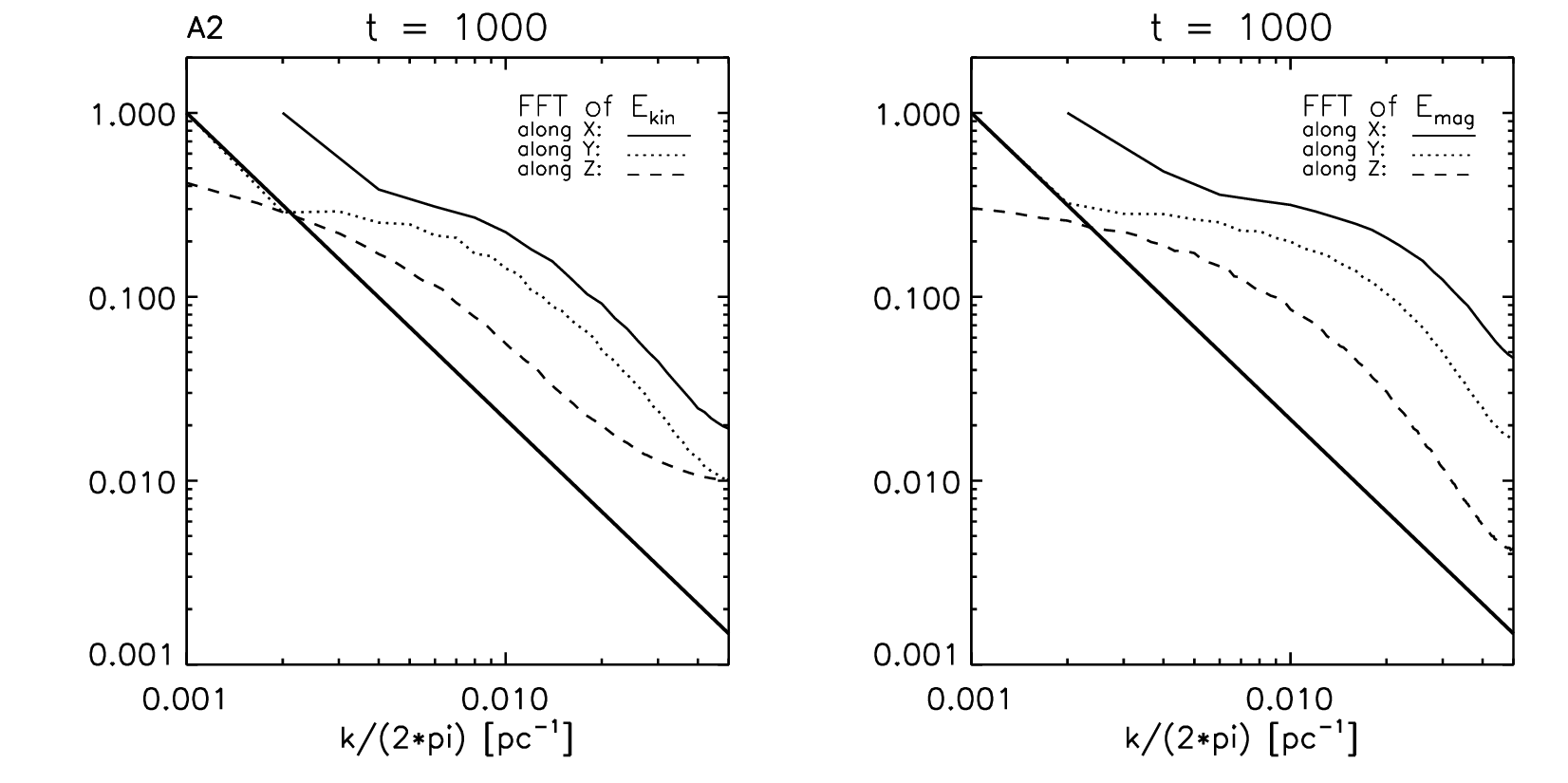}}
\centerline{\includegraphics[width=\columnwidth]{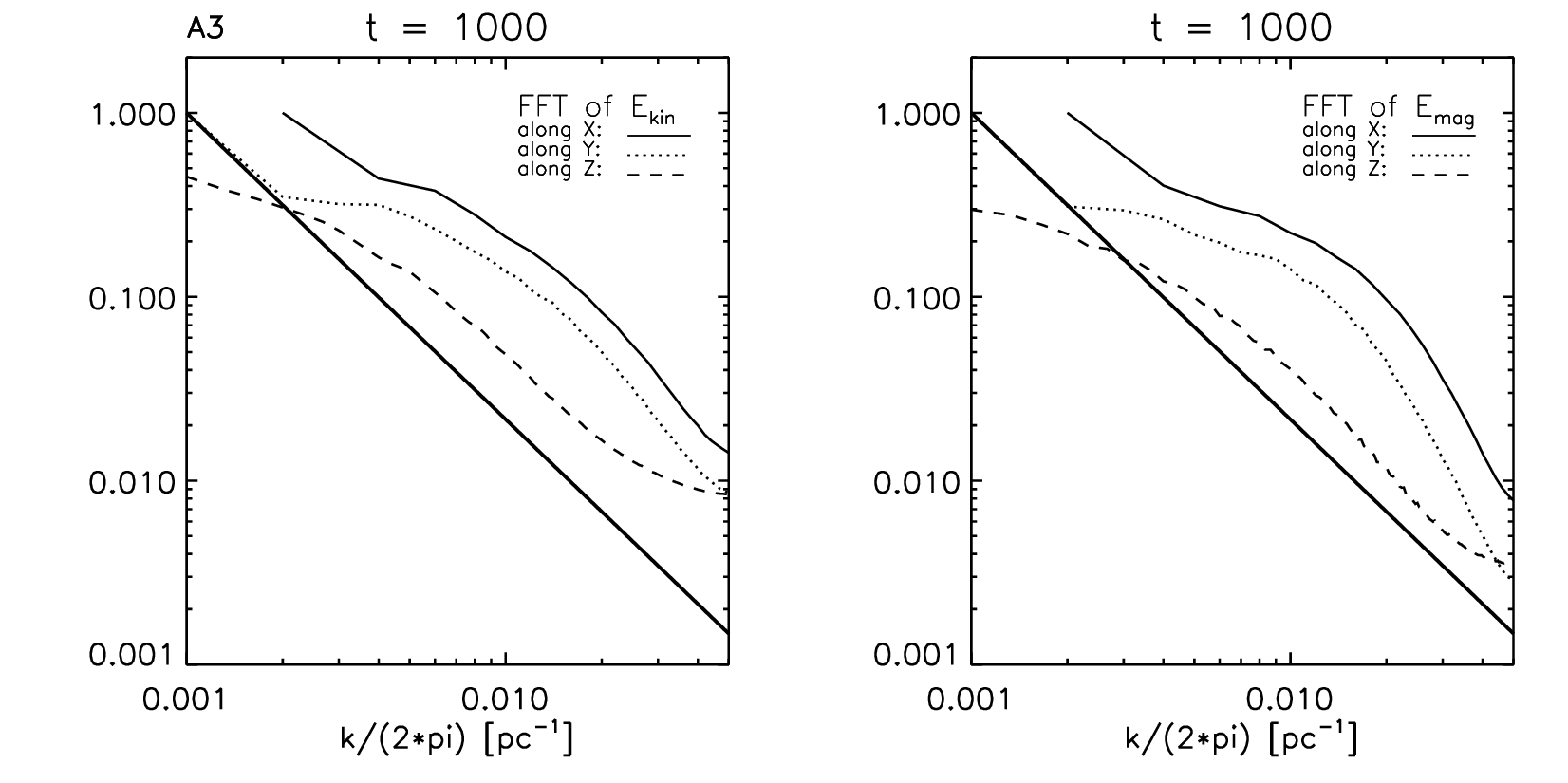}}
\centerline{\includegraphics[width=\columnwidth]{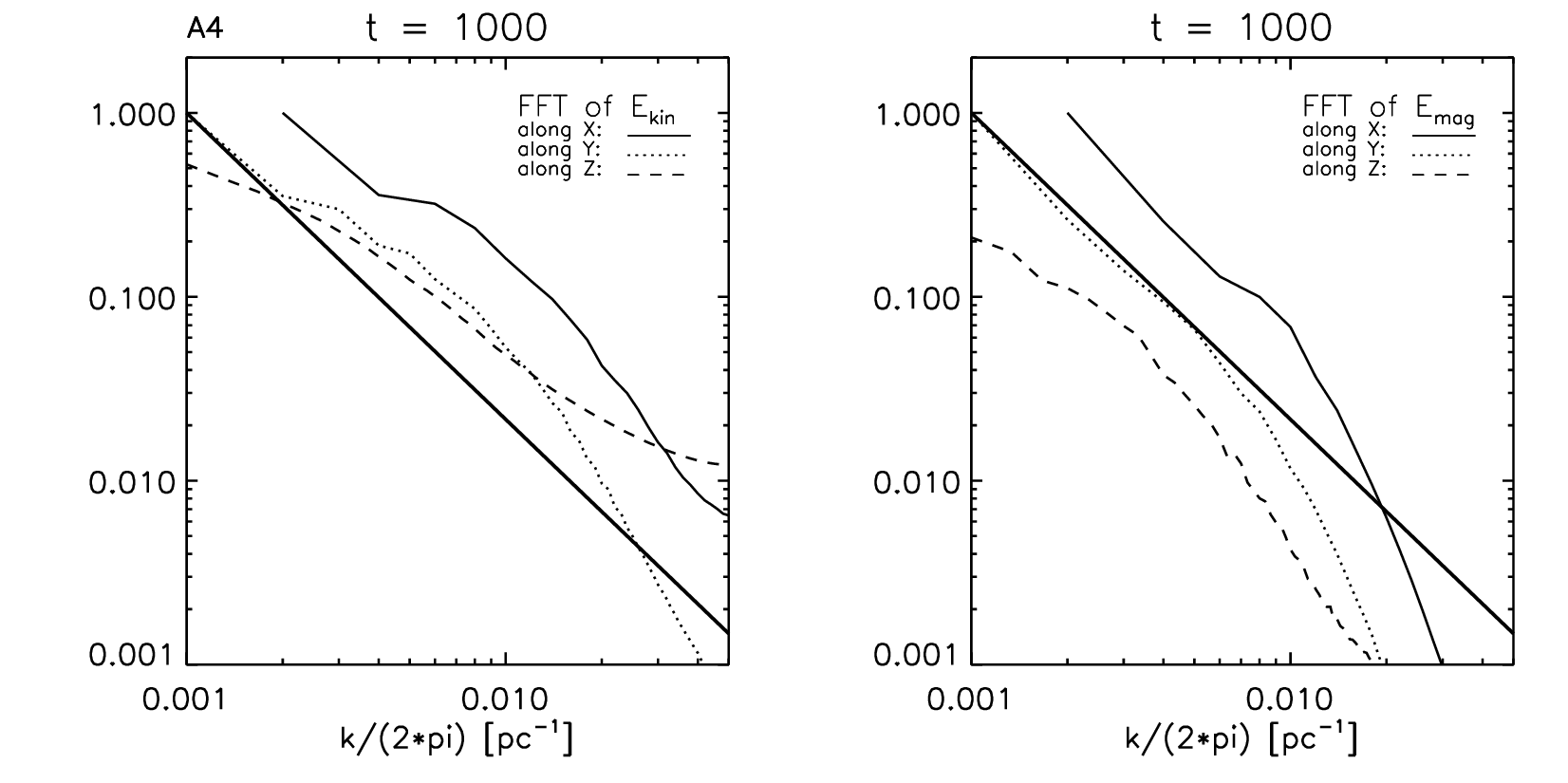}}
\centerline{\includegraphics[width=\columnwidth]{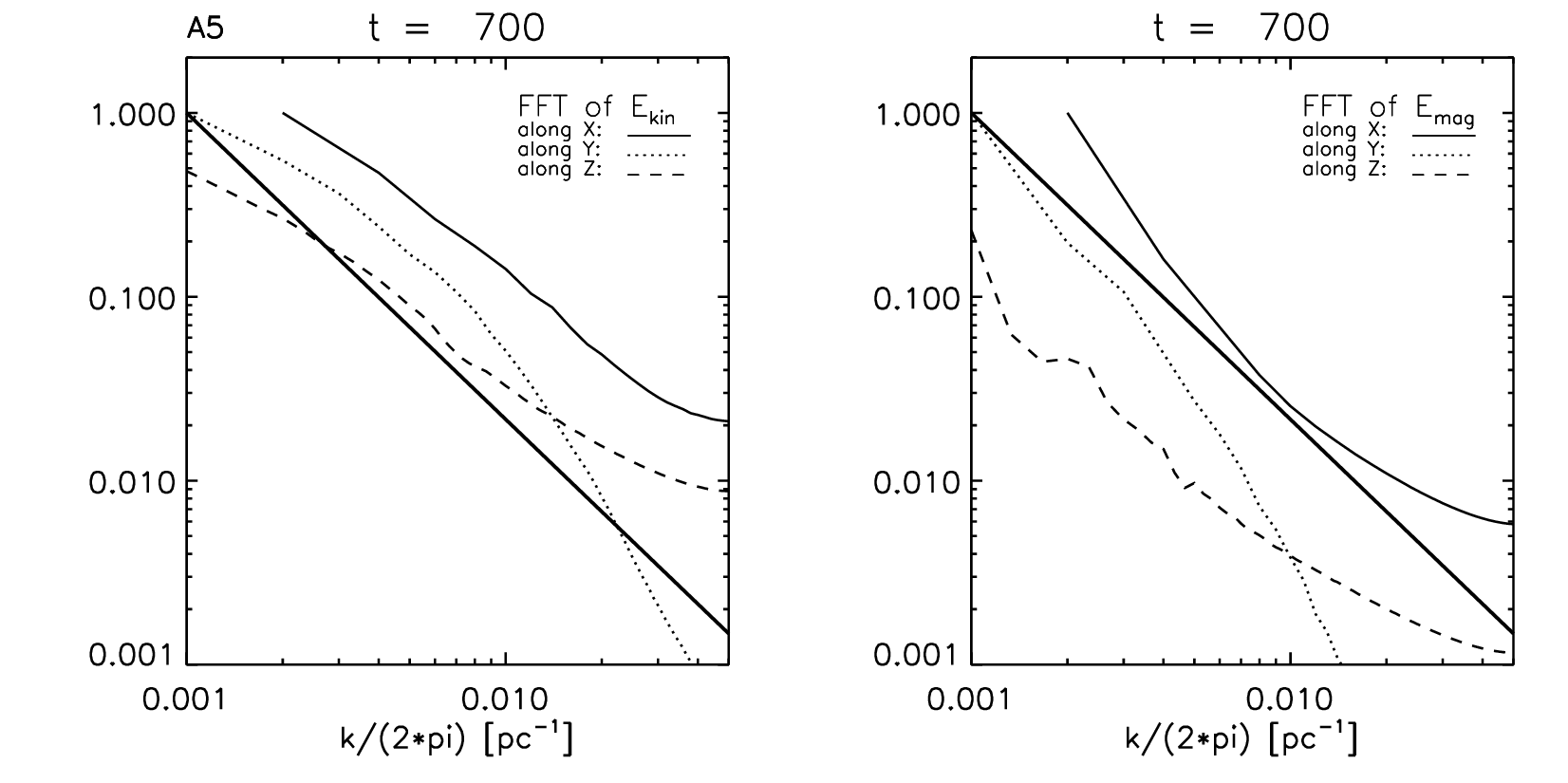}}
\caption{Kinetic (left column) and magnetic (right column) spectra computed for $t=1000\Myr$ (runs A1-A4) and $t=700\Myr$ (Run A5),
separetelly in $x$, $y$ and $z$-directions (full, dotted and dashed thin lines, respectively).
Lines representing the $k^{-5/3}$ slope (thick full lines) are shown for comparison.} 
\label{fig:fourier}
\end{figure}

To demonstrate the effect of resistivity on kinetic and magnetic turbulent
spectra we compute the Fourier transforms of kinetic and magnetic energy
densities, as in the paper by Otmianowska-Mazur et al (2007). The results are
shown in Fig~\ref{fig:fourier}. The highly anisotropic nature of turbulence is
reflected by different lines representing Fourier transforms in $x$, $y$ and
$z$-directions. We find that the kinetic spectra, which  are generally close to
the Kolmogorov spectrum $~\propto k^{-5/3}$,  depend rather weakly on magnetic
diffusivity. The large values of magnetic diffusivity $\eta = 100 \div 1000
\pc^2\Myr^{-1}$ lead only to the damping of the short-wavelength  components of
the Fourier spectrum computed in $y$-direction. 

The magnetic spectra appear to be much more sensitive to variations in magnetic
diffusivity. The plots obtained for runs A1 and A2 exhibit practically identical
spectra in all directions. This means that the diffusivity of $1 \pc^2\Myr^{-1}$
does not change the results with vanishing explicit resistivity, or in other
words the numerical resistivity of the code corresponds to the resistivity of
Run A2. The effect of resistivity is apparent through the reduced amplitude of
large-$k$ modes for $\eta= 10 \pc^2\Myr^{-1}$ (Run A3) and an apparent cutoff in
magnetic spectra around $k/2\pi \simeq 0.2 pc^{-1}$ for $\eta= 100
\pc^2\Myr^{-1}$. Further increase of magnetic diffusivity up to $\eta = 1000
\pc^2\Myr^{-1}$ leads to the signifficant reduction of amplitudes of all modes
for the Fourier tranforms performed in $z$-direction, steppening of the whole
spectrum in $z$-direction and a surprising effect of flattening of the spectrum
in $x$-direction. The latter effect may indicate a qualitative change of the
physical nature of the modes, which can plausibly attributed to the mentionned
enhancement of the exchange mode of Parker instability.

We briefly note that we neglect any small scale dynamics of the helical MHD
turbulence. This assumption allows the uniformity of our diffusion coefficients,
like the magnetic diffusivity. Detailed investigations of the influence of small
scale helical MHD turbulenceon galactic dynamos are presented by Maron \&
Blackman (2002) and Maron, Cowley and McWilliams (2004).

One can further interpret these results in terms of  topological evolution of
magnetic field which is controlled by resistivity. The topology  of magnetic
field lines determines the paths of anisotropic cosmic ray transport. For low
values of resistivity the buoyancy of cosmic rays leads to opening of magnetic
field lines through upper- and lower-$z$ boundaries. This implies that diffusive
escape of cosmic rays, along the open magnetic field lines, dominates over the
buoyancy and limits the effect of Coriolis force, which is responsible for
dynamo action.

\subsection{The effect of spiral arms \label{sect:arms}}

\begin{figure*}
\centerline{\qquad\includegraphics[width=0.45\textwidth]{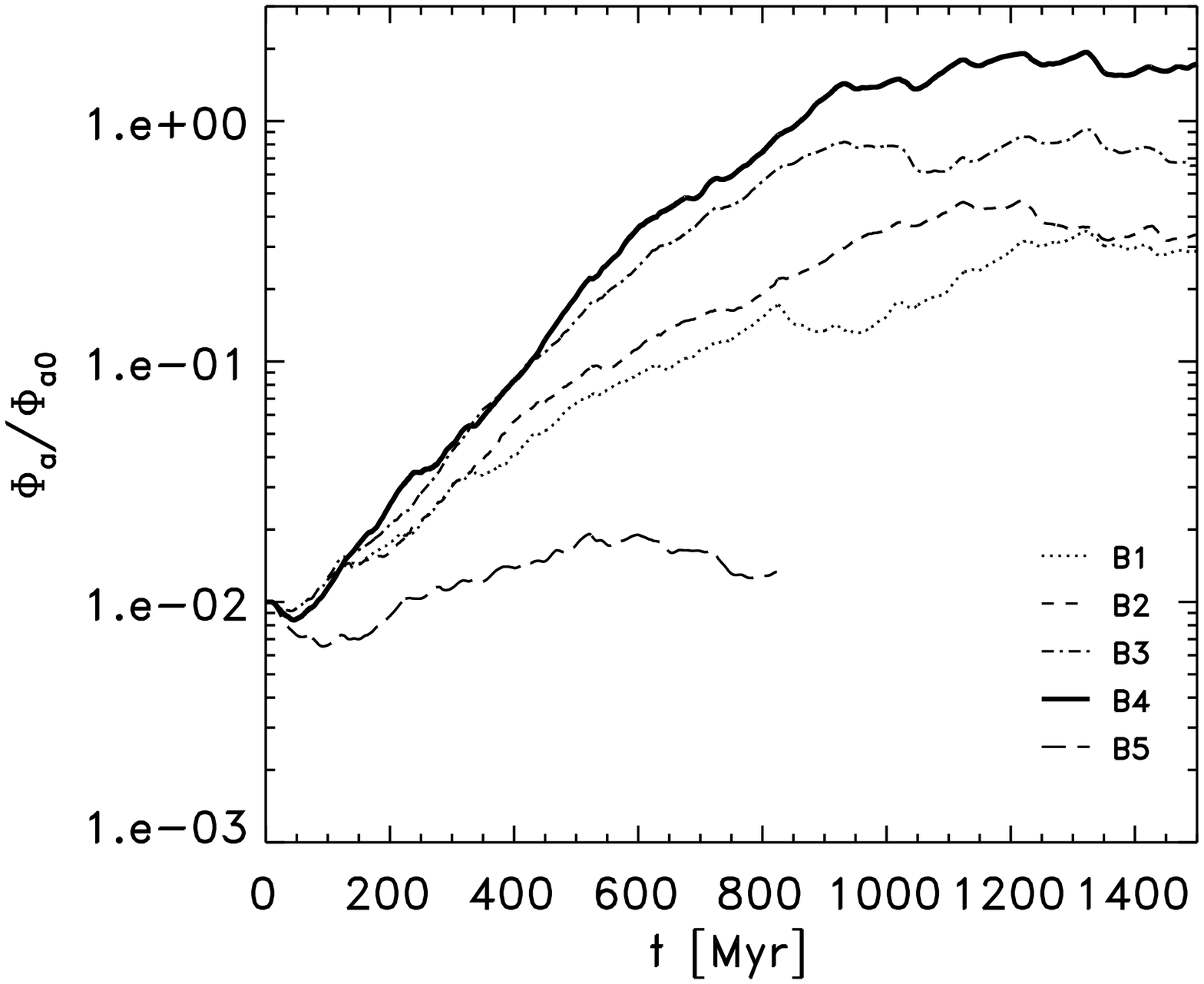}
            \qquad\includegraphics[width=0.45\textwidth]{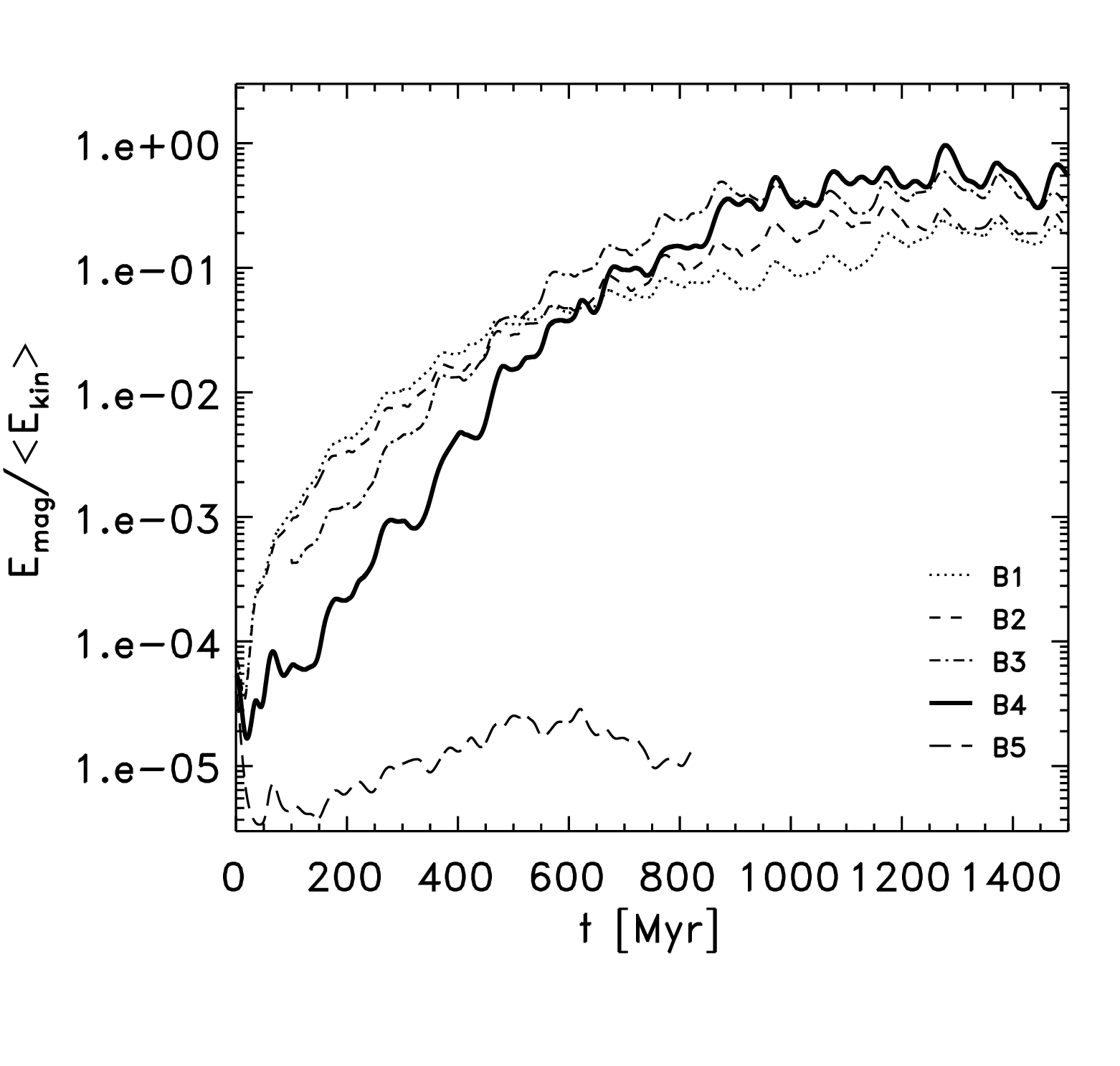}}
\caption{Time evolution of azimuthal  magnetic flux and total magnetic energy
for different values of magnetic diffusivity, in presence of temporal
modulations of SN-rate mimicking the presence of spiral arms in the simulation
series B. The curves represent respectively cases of $\eta=0$ (B1), $\eta=1$ (B2), 
$\eta=10$ (B3), $\eta=100$ (B4) and $\eta=1000$ (B5) in units $\pc^2 \Myr^{-1}$.}
\label{fig:arms}
\end{figure*}

In this section we describe the simulation series B performed for the same set
of simulation parameters as for series A with an exception that currently the
cosmic ray energy input is modulated in time by a step function. Motivation for
this kind of CR supply is the presence of spiral arms in disk galaxies. We
assume that SNe explode in arms with the rate, which is proportional to star
formation rate. We assume that arms pass trough the
volume of our local computational domain once per $100 \Myr$ and that the arm
passage takes $25 \Myr$, i.e. starting at $t=0$ we supply CRs for the first $25
\Myr$ of the $100 \Myr$  and stop CR supply for the remaining $75 \Myr$. We
enhance the SN rate in arms 4 times, so that the average SN rate over the whole
period of density wave remains the same as in the simulation series A. The
evolution of mean magnetic flux and energy is presented in Fig.~\ref{fig:arms}.

We find that in the present set of simulations, both magnetic flux and magnetic
energy grow up faster than in the case of simulations without CR modulation.
Magnetic field amplification is now  apparent even in simulations with $\eta=0$.
The only exception is the $\eta=1000 \pc^2 \Myr^{-1}$ simulation which does not show a
noticeable amplification of magnetic field, even in the presence of CR
modulation. This indicates that the temporal modulation of SN rate acts in the
same way as increasing magnetic diffusivity in the range of $\eta=0 \div 100
\pc^2\Myr^{-1}$.

To interpret the above results we suggest  the following scenario: CRs supplied
to the system trigger Parker instability and leave the disk volume via combined
buoyant and diffusive transport. Vertical magnetic loops form efficiently during
the period of enhanced SN activity, but later on, in absence of CR
perturbations, magnetic field tends to relax before the next spiral arm passage.
In absence of CR forcing, in the inter-arm regions even a small resistivity is
sufficient to relax magnetic field structure to the horizontal more regular
configuration which suppresses excessive losses of CRs via the diffusive transport.
Thus the efficiency of magnetic field amplification is enhanced.

\subsection{Dependence of magnetic field amplification on SN-rate}

\begin{figure*}
\centerline{\qquad\includegraphics[width=0.45\textwidth]{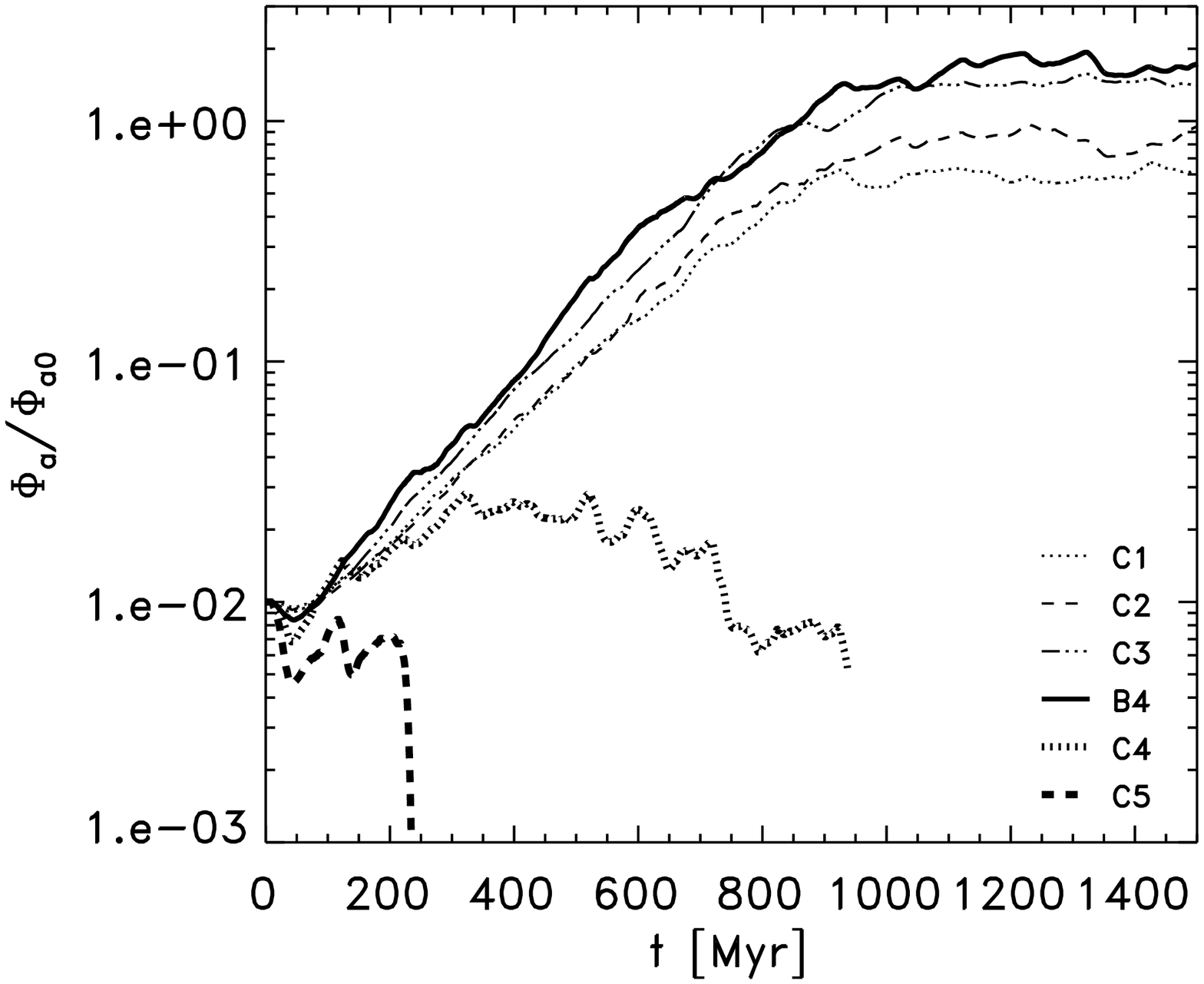}
            \qquad\includegraphics[width=0.45\textwidth]{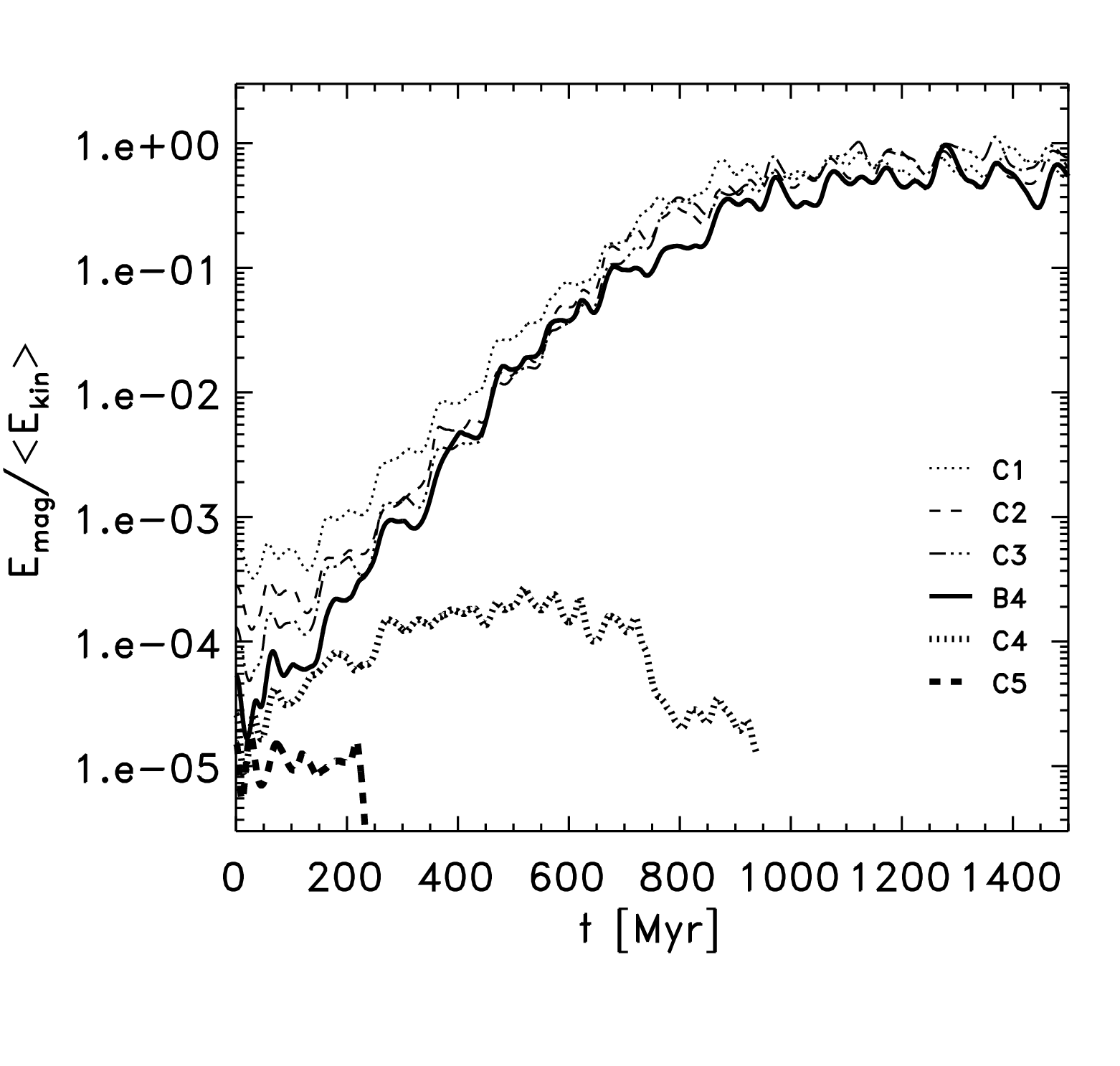}}
\caption{Time evolution of azimuthal  magnetic flux and total magnetic energy
for different values of SN rate applied in simulation series C, together with run B4, in the presence
of temporal modulations of SN-rate. Line assignments are respectively: $f_{SN}=15$ (C1), 
$f_{SN}=30$ (C2), $f_{SN}=60$ (C3), $f_{SN}=130$ (B4), $f_{SN}=250$ (C4), 
and $f_{SN}=500$ (C5) supernova explosions per squared kpc per Myr.}
\label{fig:fsn}
\end{figure*}

In all runs of the simulation series A and B we adopted the fiducial value of
$f_{SN} = 130 \kpc^{-2} \Myr^{-1}$ derived from the global model of Milky Way by
Ferriere (1998) at galactocentric radius $R_G = 5 \kpc$.  In this section we
describe simulation series C performed for different SN rates  $f_{SN} = 15, 30,
60, 250$ and $500\kpc^{-2} \Myr^{-1} $, together with run B4 
($f_{SN}=130\kpc^{-2} \Myr^{-1} $), to examine the effect of SN rate on
magnetic field amplification.  The results are shown in Fig.~\ref{fig:fsn}. In
all cases the SN input is modulated in a manner described in
Sect.~\ref{sect:arms}.

\begin{figure}
\centerline{\includegraphics[width=0.45\textwidth]{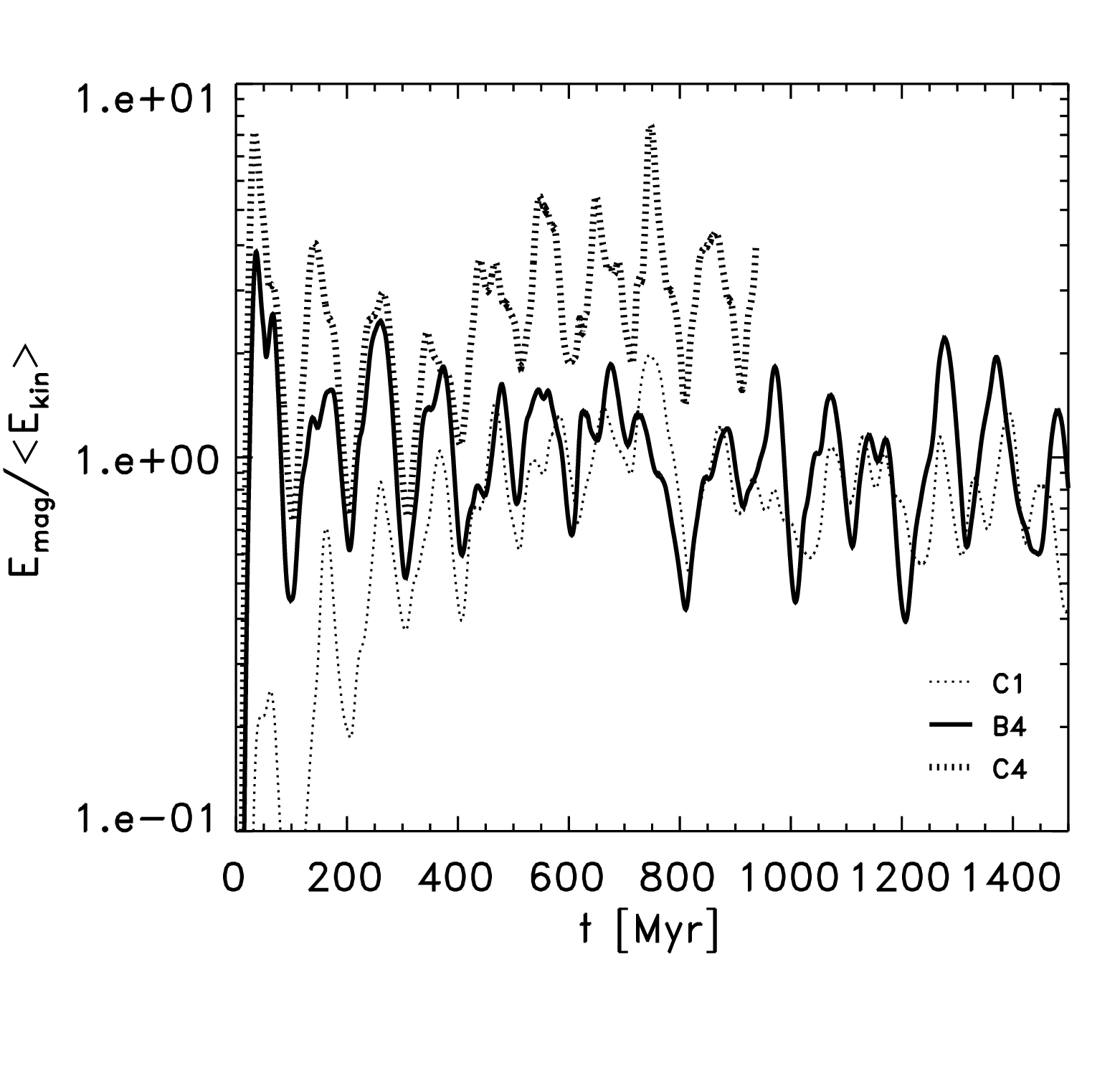}}
\caption{Time evolution of the ratio of energies of vertical to horizontal
magnetic field components for different (modulated) SN rates $f_{SN}=15$ (C1), $f_{SN}=130$ (B4)
and $f_{SN}=250$ (C4) supernova explosions per squared kpc per Myr.}
\label{fig:bz-by-fsn}
\end{figure}

One can notice that the magnetic field amplification rates and the final
saturation levels of both magnetic flux and magnetic energy grow with $f_{SN}$
as long as the SN rate is lower or equal to the fiducial, realistic  value of
$f_{SN} = 130 \kpc^{-2} \,\Myr^{-1}$ at galactocentric radius $R_G=5\kpc$, when
all other disk parameters are fixed. The e-folding times of magnetic flux
deduced from the left panel of Fig.~\ref{fig:fsn} are respectively 150 Myr for
$f_{SN} = 130 \kpc^{-2} \,\Myr^{-1}$ (Run B4) and 190 Myr for  $f_{SN} = 15
\kpc^{-2} \,\Myr^{-1}$ (Run C1). We note therefore that the magnetic field
amplification rate is relatively insensitive to the magnitude of SN rate within
the range of SN rates spanning one decade below the fiducial value.  We note
also that magnetic field amplification saturates at the level of equipartition
of magnetic and kinetic energies in the case of those simulation runs of series C for which
the amplification holds.

When the SN rate is doubled, with respect to the fiducial value, then only a short period of magnetic field
amplification is observed, until $t = 500 \,\Myr$, and if SN rate is doubled
once again then magnetic field decays. The above results show that
magnetic field amplification holds in a wide range of SN rates and that the
realistic values of SN rates are optimal for the galactic dynamo process.
Similarly as in Sect.~\ref{sect:eta} we show the ratio of energies in vertical
to horizontal magnetic field components in Fig.~\ref{fig:bz-by-fsn}. We find
that for SN rates up to $f_{SN} = 130 \kpc^{-2} \,\Myr^{-1}$ the efficient
magnetic field amplification is associated with the ratio of vertical to
horizontal magnetic field energies fluctuating around one, and in the case of
excessive CR supply ($f_{SN} = 250 \kpc^{-2}\, \Myr^{-1}$ and more) the vertical
magnetic field energy dominates and magnetic field ceases to grow.

\subsection{Dependence of magnetic field amplification the grid resolution \label{sect:resol}}

\begin{figure*}
\centerline{\qquad\includegraphics[width=0.45\textwidth]{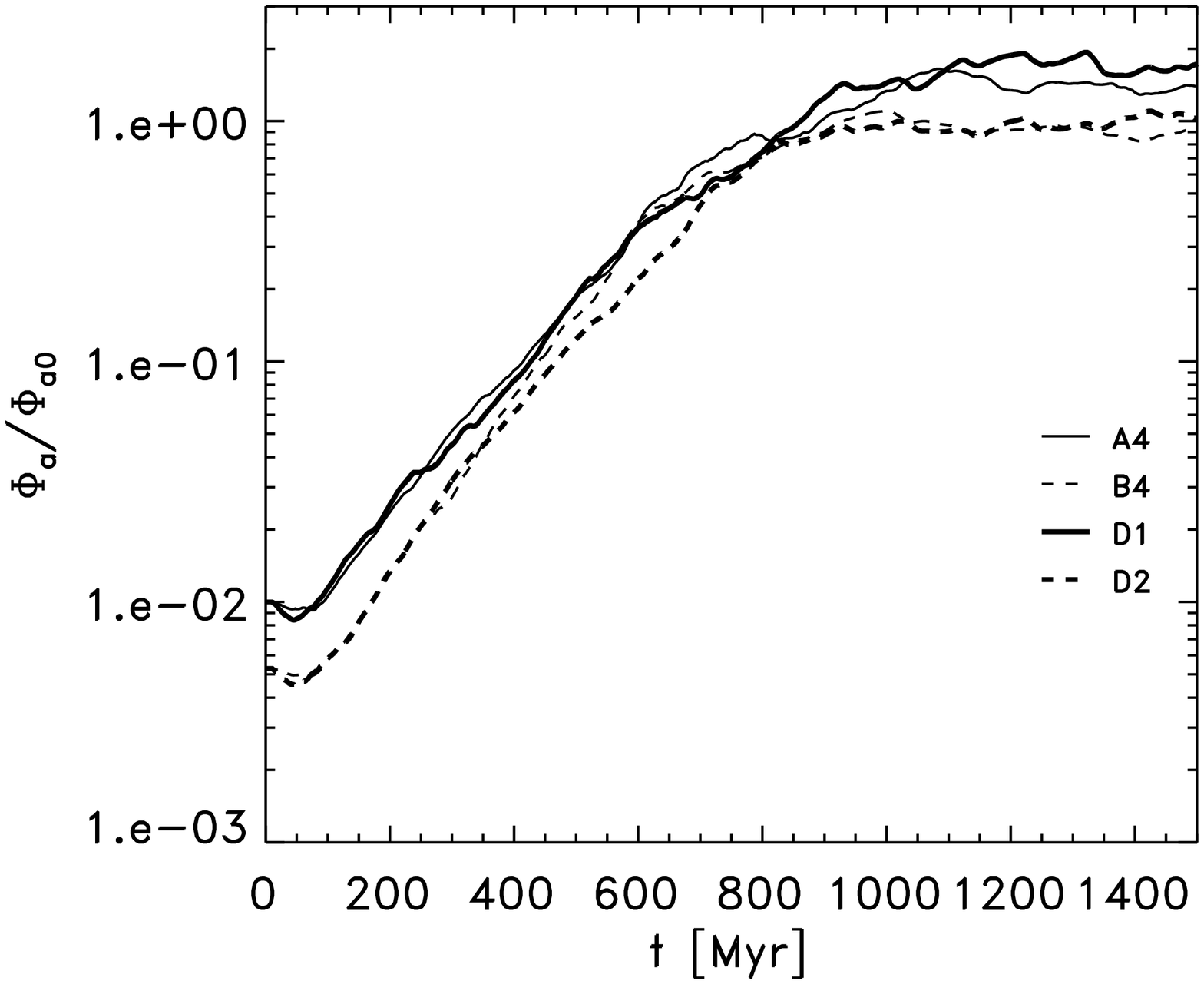}
            \qquad\includegraphics[width=0.45\textwidth]{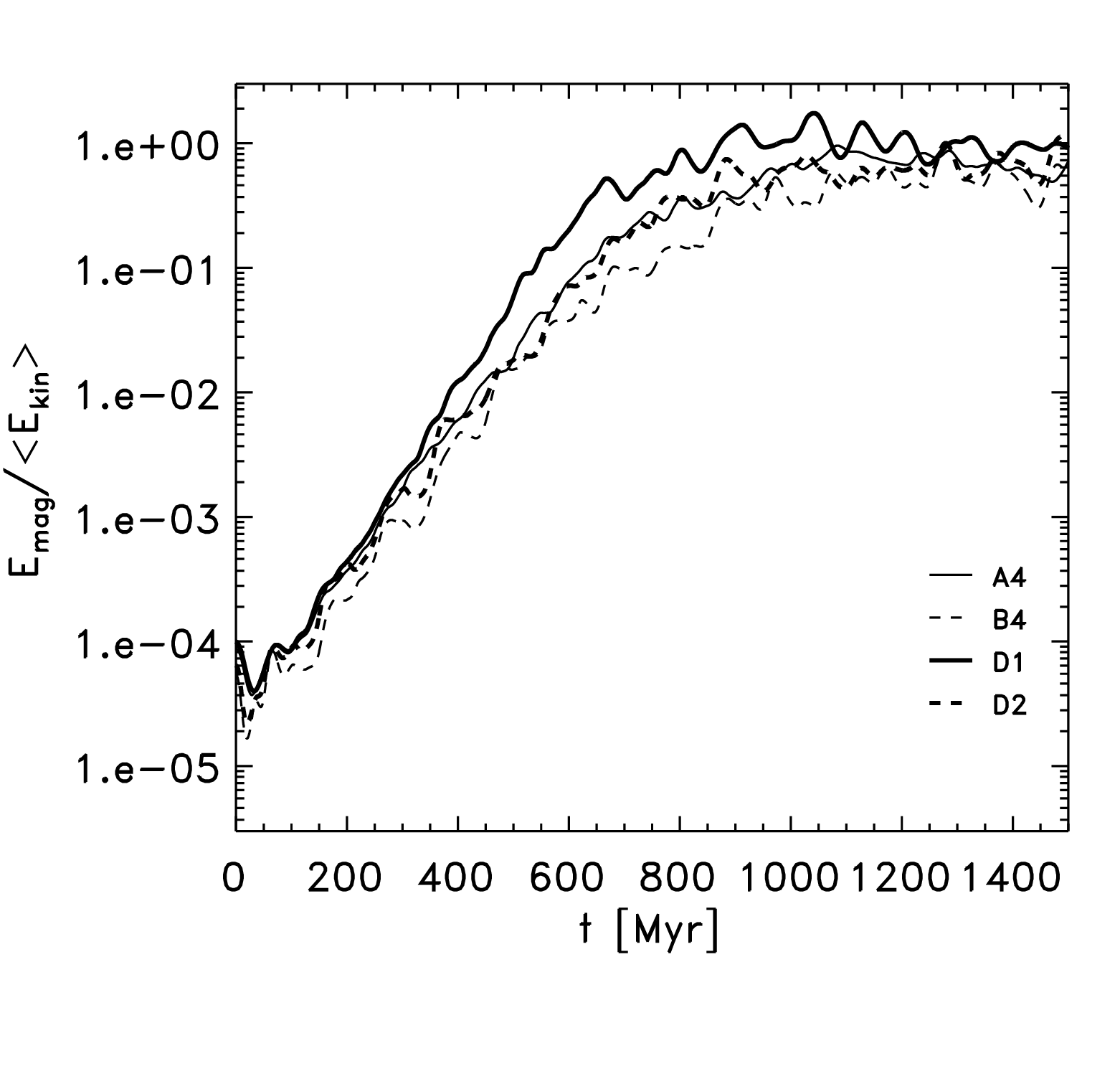}}
\caption{Time evolution of the azimuthal magnetic flux and the total magnetic
energy for simulations with grid resolutions $(10 \pc)^3$ (runs A4 and B4) and
$(20 \pc)^3$ (runs D1 and D2). }
\label{resol}
\end{figure*}

In order to check the influence of the grid resolution on simulation results we
increase the cell size to $(20\pc)^3$ in simulations D1 and D2 and apply the
same parameters as in simulations A4 and B4, respectively. In
Fig.~\ref{resol} we show the evolution of the total flux of the azimuthal
magnetic field component and the total magnetic energy for simulations D1 and D2
together with analogous curves for simulations A4 and B4, shown previously in
Figs.~\ref{fig:eta} and \ref{fig:arms}. It is apparent that the results obtained
at both resolutions are very similar, although a slightly faster growth of
magnetic field is observed in simulations performed at the lower resolution, 
which can be explained by somewhat larger numerical resistivity. 

\subsection{Dependence of magnetic field amplification on CR diffusion
coefficients \label{sect:difco}}

The aim of simulation series E is to examine the effect of variations of the CR
diffusion coefficients on magnetic field amplification. All simulations of series
E are performed with the resolution $(20\pc)^3$.  The reduced grid resolution
makes it possible to enlarge CR diffusion coefficients to realistic values, 
while preserving acceptable timesteps for the explicit integration algorithm of
the CR diffussion-advection equation.

\begin{figure*}
\centerline{\qquad\includegraphics[width=0.45\textwidth]{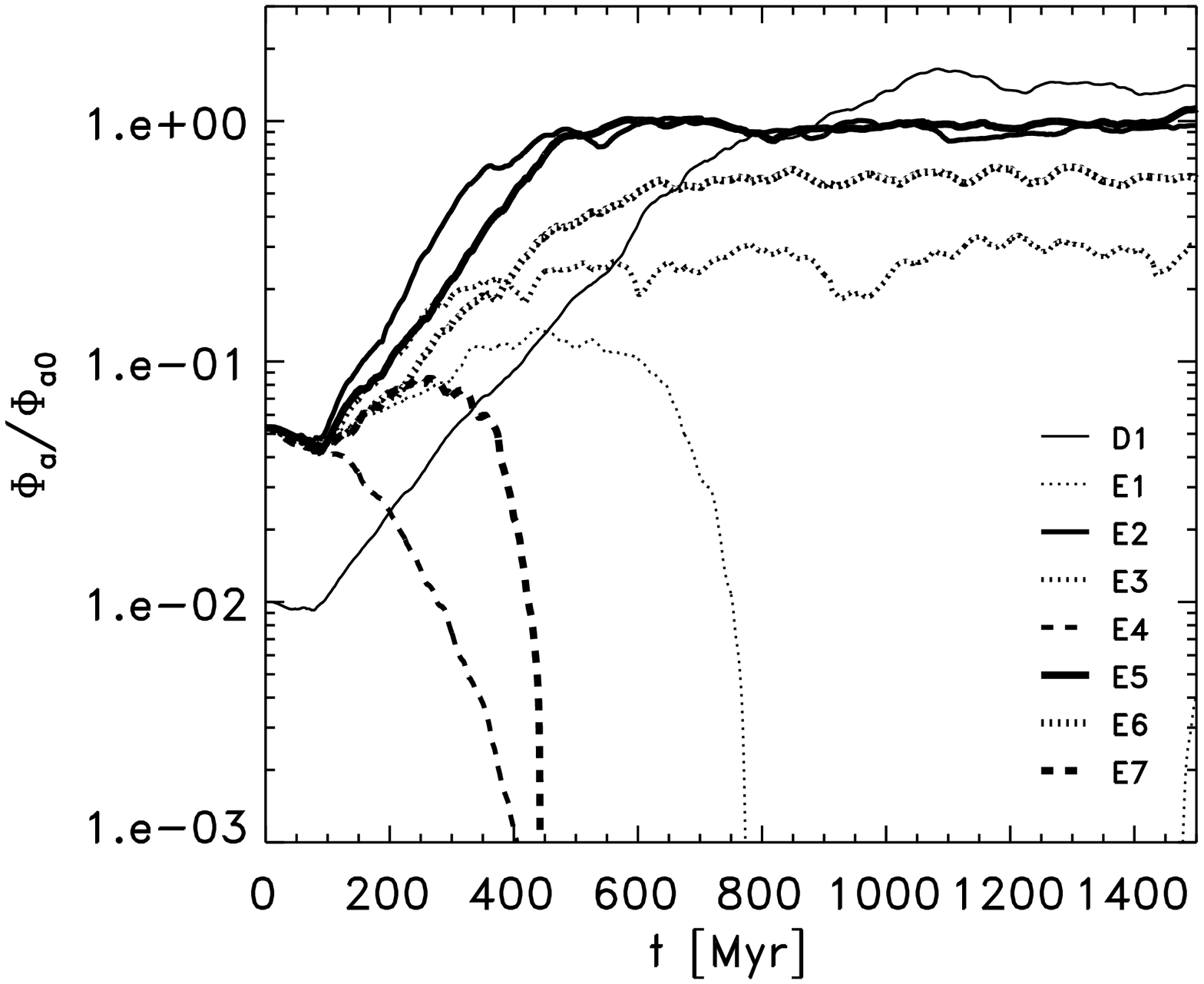}
            \qquad\includegraphics[width=0.45\textwidth]{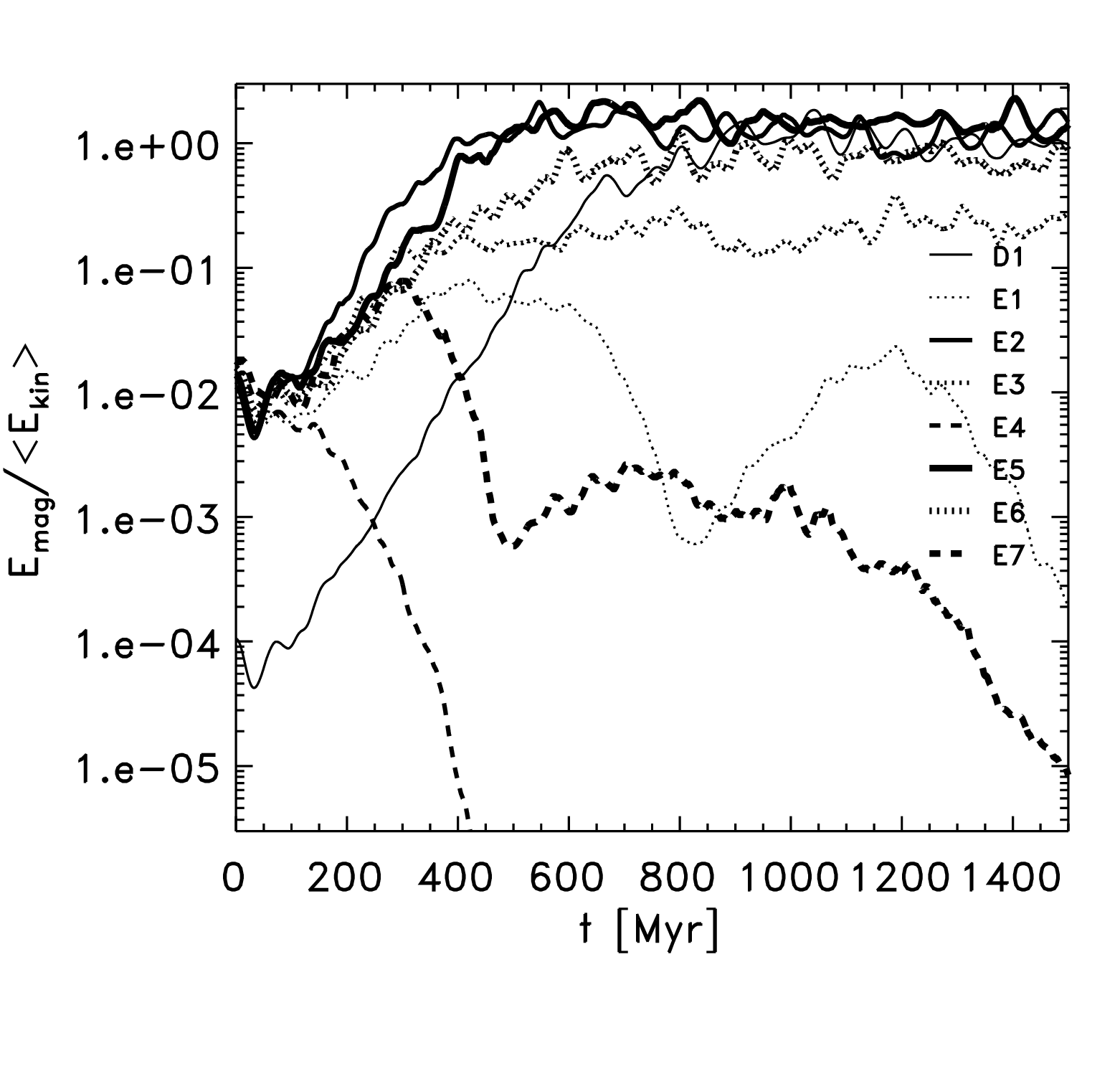}}
\caption{Time evolution of the azimuthal  magnetic flux and the total magnetic
energy for different values of the parallel and perpendicular CR diffusion
coefficients. Thin lines are used for 
 $K_\paral=1\times10^4 $ (runs D1 and E1), 
 mid lines are used for  $K_\paral=3\times10^4 $ (runs E2 E3 and E4)
  and thick lines are used for $K_\paral=1\times10^5 $ (runs E5 E6 and E7).
Full lines denote $K_\perp=1\times10^3 $ (runs D1, E2 and E5), 
dotted lines denote $K_\perp=3\times10^3 $ (runs E1, E3 and E6), 
dashed lines  $K_\perp=1\times10^4 $ (runs E4 and E7). All diffusion coefficients given in units 
$\pc^2\Myr^{-1}$.}
\label{fig:cr-diff}
\end{figure*}

In the present simulation series E we vary parallel and perpendicular diffusion
coefficients choosing different pairs from the set: $K_\paral =1\times 10^4$, 
$3\times 10^4$ and $1\times 10^5  \pc^2 \Myr^{-1}$, and  $K_\perp = 1\times
10^3$, $3\times 10^3$ and $1\times 10^4 \pc^2 \Myr^{-1}$. The results of new
simulations as compared to the run D1 are presented in Fig.~\ref{fig:cr-diff}.

\begin{figure}
\centerline{\qquad\includegraphics[width=0.45\textwidth]{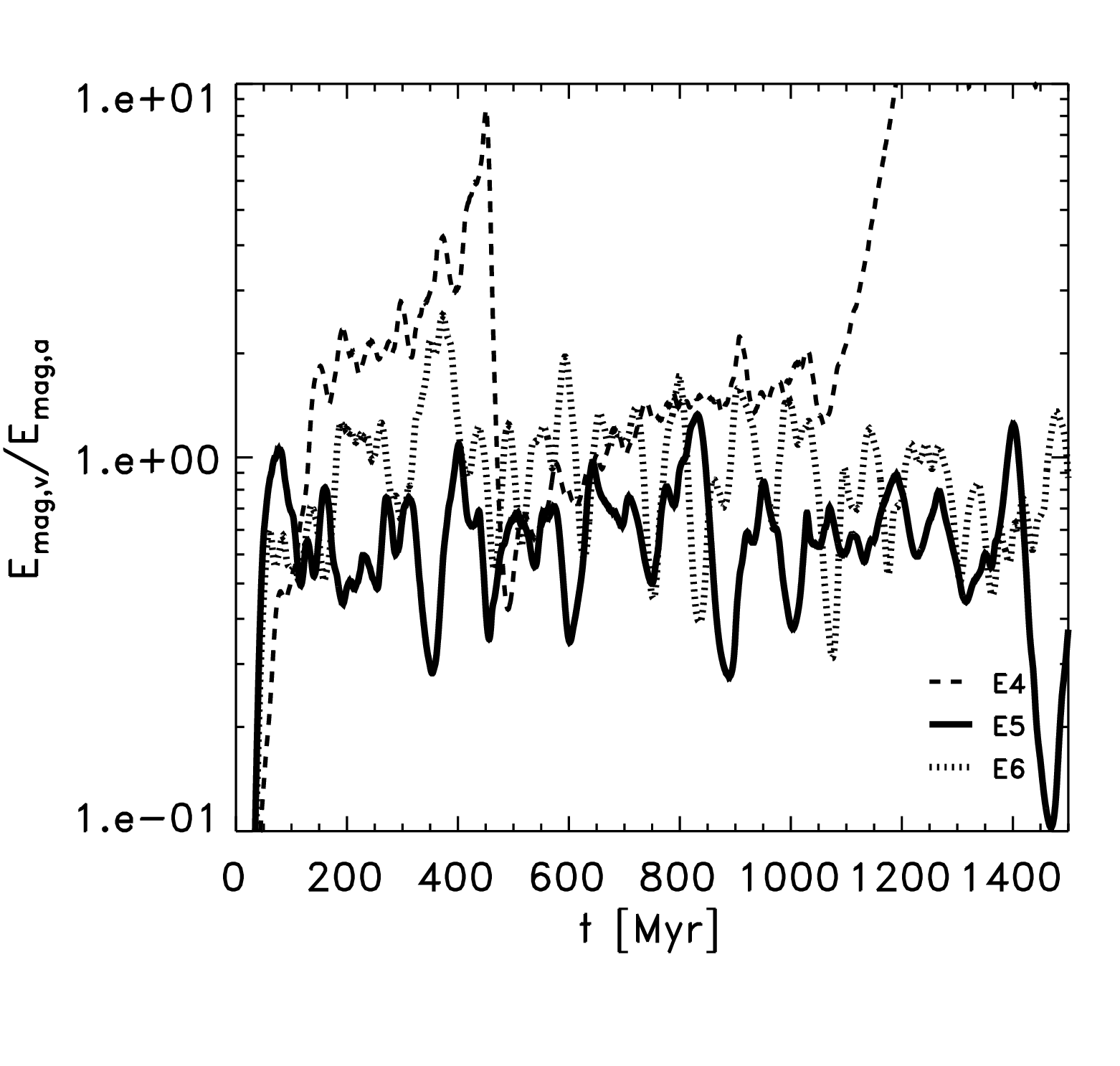}}
\caption{Time evolution of the ratio of energies of vertical to horizontal
magnetic field components for different values of the perpendicular CR diffusion
coefficients  $K_\paral=3\times10^4$ and $K_\perp=1\times10^4
$ (E4), $K_\paral=1\times10^5 $ and
$K_\perp=1\times10^3$ (E5) and $K_\paral=1\times10^5
$, $K_\perp=3\times10^3 $ (E6). All diffusion coefficients Gaven in units 
$\pc^2\Myr^{-1}$. }
\label{bz-by-kperp}
\end{figure}

The results of simulation series E can be summarized as follows. We
note, that magnetic field growth rate and the saturation values of magnetic flux
and energy depend particularly on the choice of $K_\paral$ and $K_\perp$. When
$K_\paral$ is increased 3 and 10 times with respect to D1, the initial growth of
magnetic field becomes slightly faster and the saturation level becomes lower by
a factor of $2\div3$, provided that $K_\perp$ is not too large. For $K_\paral =
3\times 10^4\pc^2 \Myr^{-1}$ amplification holds for $K_\perp = 1\times 10^3
\pc^2 \Myr^{-1}$ and $3\times 10^3\pc^2 \Myr^{-1}$, but for  $K_\perp = 1\times
10^4$ magnetic field decays. Similarly, for $K_\paral = 1\times 10^5\pc^2
\Myr^{-1}$ amplification holds for $K_\perp = 1\times 10^3\pc^2 \Myr^{-1}$ and
$K_\perp = 3\times 10^3\pc^2 \Myr^{-1}$, but for  $K_\perp = 1\times 10^4\pc^2
\Myr^{-1}$ we find only initial growth until $t=450 \ \Myr$ and decay
thereafter.  The present results indicate that magnetic field amplification is
possible only for $K_\perp \leq  3\times 10^3\pc^2 \Myr^{-1} \simeq  10^{27}
\cm^2 s^{-1} $. {\em Therefore the anisotropy of CR diffusion seems to be a crucial
condition for magnetic field amplification in the process of the CR-driven
dynamo.}

In the subsequent Fig.~\ref{bz-by-kperp} we show the energy ratio of vertical to
azimuthal magnetic field components  for  runs E4, E5 and E6, corresponding to
three different values of $K_\perp$ and  $K_\paral = 1\times
10^5\pc^2\Myr^{-1}$. Comparing Figs.~\ref{fig:cr-diff} and \ref{bz-by-kperp} we find
that in the case of two simulation runs E4 and E5 (two smaller values of
$K_\perp$) the energy ratio of vertical to azimuthal magnetic field components
varies in the range $\sim 0.3 \div 2$, corresponding to efficient growth
magnetic energy. For the largest value of $K_\perp=10^4 $ the  magnetic energy
ratio becomes occasionally larger by an order of magnitude, magnetic field
decays.

\subsection{The issue of energy equipartition}

\begin{figure}
\centerline{\qquad\includegraphics[width=0.45\textwidth]{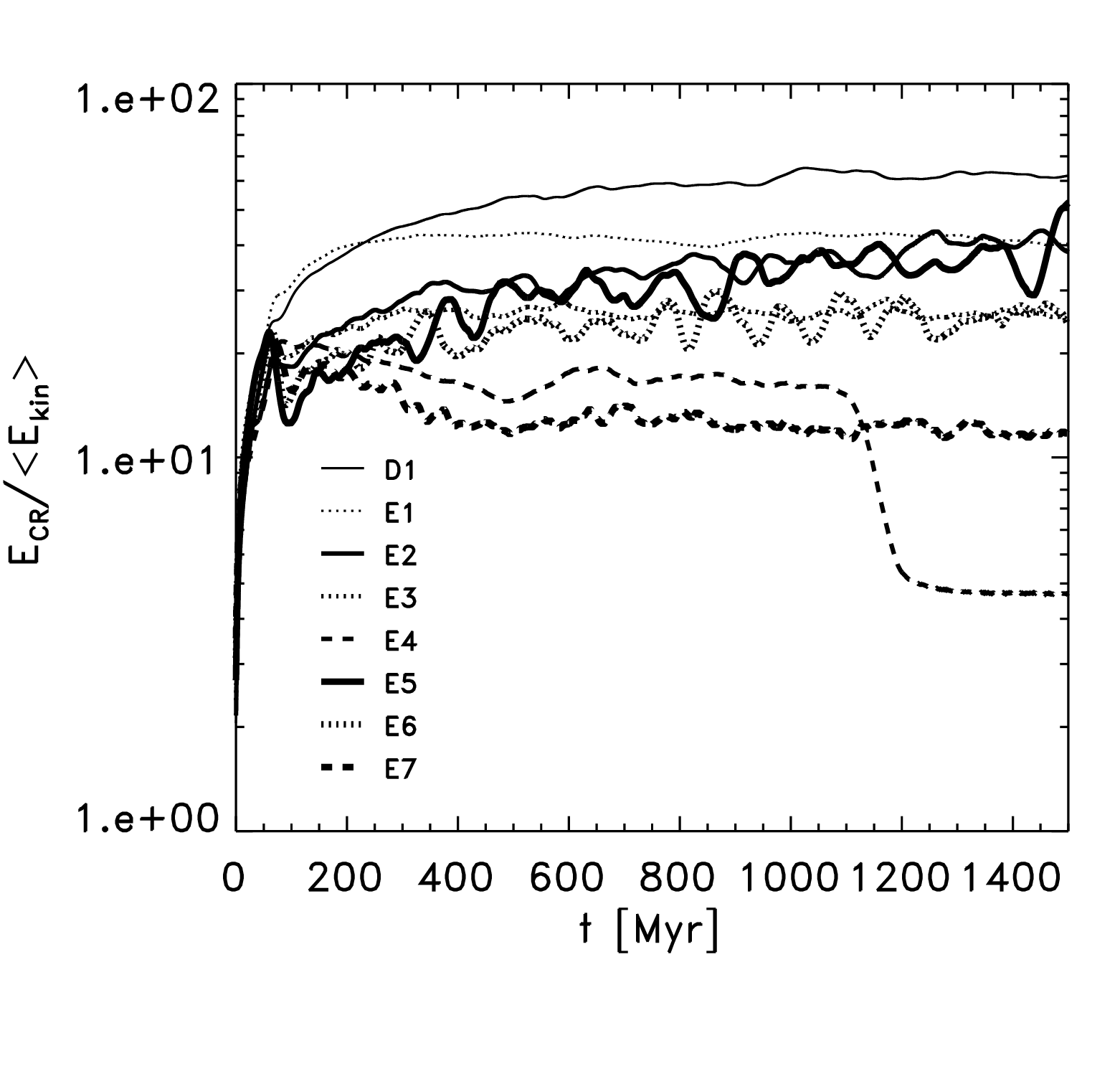}}
\caption{Time evolution of the ratio of CR to time-averaged kinetic energy for
different values CR diffusion coefficients in simulation series E. Line
assignments are the same as in Fig.~\ref{fig:cr-diff}}
\label{fig:crtokin}
\end{figure}

The results presented so far demonstrate that magnetic fields amplified by the
CR-driven dynamo saturate near equipartition of magnetic and gas kinetic
energies. It is commonly expected that CRs remain in energetic equipartition
with gas and magnetic field as well.

To investigate the relation of CR to other forms of energies, we plot in
Fig.~\ref{fig:crtokin} the time evolution of the ratio of CR to the
time-averaged turbulent kinetic energy (we subtract the kinetic energy of the
large-scale shear flow) for different values of CR diffusion coefficients (the
simulation series E).  We find that, depending on diffusion coefficients, CR
energy is larger than the turbulent kinetic energy by a factor of $10-50$, while
the magnetic field energy, according to the results presented in
Fig.~\ref{fig:cr-diff}, is close to the gas turbulent energy.

It is apparent that for the first few hundred Myr CRs accumulate quickly in the
disk since they are trapped by horizontal magnetic field. The ratio of CR to
kinetic energies saturates as soon as vertical magnetic field component appears,
due to buoyancy, enabling diffusive transport of CRs out of the disk.

When the cosmic ray diffusion coefficients are larger, one can find that the
ratio of CR to kinetic energies is lower. We note, that cosmic rays find an
easier way to leave the disk, when the parallel and perpendicular diffusion
coefficients are larger. The results shown in Fig.~\ref{fig:crtokin} indicate
that the ratio of CR to kinetic energies anticorrelates with both: the parallel
and perpendicular CR diffusion coefficients. Due to the mentioned timestep
limitation, in the simulations presented in this paper we could only adopt the
values of $K_\paral$ reaching at most $3 \times 10^{28} \cm^2\s^{-1}$, that are
still smaller than $10^{29} \cm^2\s^{-1}$ mentioned by other authors (e.g.
Jokipii 1999). Although the tendency of lowering the CR energy with the
magnitude of the parallel CR diffusion coefficient seems promising, one should
not expect that values as large as  $K_\paral \simeq 10^{29} \cm^2\s^{-1}$ will 
reduce the problem of CR energy excess.

Another factor, which may significantly influence the relation between CR and
other form of energies, is the choice of periodic and shear-periodic boundary
conditions in the horizontal directions of the computational box.  In real
galactic disks, CR diffusion is expected along horizontal magnetic field lines.
In the case of periodic-type boundary conditions, CRs are trapped in the disk
volume by predominantly horizontal magnetic field. This kind of trapping can be
released, however, only in the global galactic disk simulations.

\section{Summary and conclusions \label{sect:summary}}

In the present paper we described an extensive series of simulations and
presented parameter study of the CR-driven dynamo in a galaxy, characterized by
the essential parameters typical for the Milky Way at galactocentric radius
$R_G=5\kpc$. In the presented study the magnetic diffusivity, as well as the
parallel and perpendicular CR diffusion coefficients have been considered as
free parameters, and dedicated simulation series have been performed to
investigate their influence on the efficiency of the CR-driven dynamo process.
The results of the parameter study can be summarized as follows:

(1) The magnitude of magnetic diffusivity influences essentially the efficiency
of magnetic field amplification. The most favorable value of magnetic
diffusivity is $100 \pc^2\Myr^{-1}  \simeq 3 \times 10^{25} \cm^2\s^{-1}$, the
value comparable, although lower than the the value of turbulent diffusivity of
ISM deduced from observational data.

(2) The efficiency of magnetic field amplification is enhanced by temporal
modulation of the CR supply. An effect of this kind may be associated with the
periodicity of star formation and supernova activity induced by spiral arms. The
enhancement is apparent at lower values of magnetic diffusivity and is less
significant at the optimal value of $\eta = 100 \pc^2\Myr^{-1}$.

(3) Magnetic field amplification rate is relatively insensitive to the magnitude
of SN rate within the range of SN rates spanning one decade below the value
$f_{SN} = 130 \kpc^{-2} \,\Myr^{-1}$ typical for galactocentric radius
$R_G=5\kpc$. We note also that magnetic field amplification saturates at the
level of equipartition of magnetic and kinetic energies for all supernova rates
for which amplification holds. Magnetic field is no longer amplified, if SN rate
is further enhanced by factors 2 and 4 with respect to the realistic value,
while other quantities (like e.g. gas column density) remain fixed.

(4) Magnetic field amplification in the CR-driven dynamo relies on anisotropic
diffusion of cosmic rays. From the limited set of simulations of series E, one
can deduce that the magnetic field amplification is possible only for $K_\perp
\leq 3\times 10^3\pc^2 \Myr^{-1} \simeq  10^{27} \cm^2 s^{-1} $, and for all
considered values of the parallel diffusion coefficient $K_\paral$ in the range
$3\div30\times 10^{27} \cm^2 s^{-1}$. Therefore, the 5\% ratio of the
perpendicular to parallel diffusion coefficients postulated by Giaccalone \&
Jokipii (1998) falls within the amplification range.

(5) By varying various parameters: magnetic diffusivity, supernova rate and the
CR diffusion coefficients, we have found that the favorable conditions for
magnetic field amplification correspond to approximately equal energies of the
vertical and azimuthal magnetic field components in the case of buoyancy driven
dynamo. An excess or deficit of vertical magnetic field with respect to the
azimuthal one corresponds to significantly less efficient amplification or even
decay of magnetic field.

(6) We note the problem indicated previously by Snodin et al (2005), that in all
simulations the CR energy in the computational domain exceeds the turbulent
kinetic energy and magnetic energy by more than one order of magnitudes.
Moreover, the lowest ratios of CR to kinetic energies emerge for the largest
values of the parallel diffusion coefficient.  It seems not plausible, however,
that an enlargement of diffusion coefficients up to fully realistic values will
reduce the excess of cosmic rays energy in the disk. It seems also, that the
ratio of CR to other forms of energy in the ISM is not yet well restricted on
observational grounds (Strong et al 2007). On the other hand, the currently used
shearing box approximation does not permit CRs to leave the disk by means of
diffusion along the predominantly horizontal magnetic field. Therefore, we
suggest that subsequent work on the CR-driven dynamo, aimed to solve this
problem, should be done in the framework of global galactic disk simulations.


\acknowledgements
This work was supported from the Polish Committee for Scientific Research (KBN)
through the grants PB 0656/P03D/2004/26 and 2693/H03/2006/31.


\begin{thebibliography}{}

\bibitem[Beck 2007]{beck-07}
   Beck R., 2007, arXiv:0711.4700
\bibitem[Beck \& Krause 2005]{beck-06}
   Beck, R, Krause, M., 2005, Astronomische Nachrichten 326
\bibitem[Berezinski et al. 1990]{berezinski-etal-90}
   Berezinskii, V.S., Bulanov, S.V., Dogiel, V.A., Ginzburg, V.L., Ptuskin, V.S.,
   {\em Astrophysics of cosmic rays}, Amsterdam: North-Holland, 1990.
\bibitem[Chi \& Wolfendale 1993]{chi-wolfendale-93} 
  Chi, X., Wolfendale, A.W., Nature 362, 610 (1993)
\bibitem[Ferriere 1998]{ferriere-98}
   Ferriere, K. 1998, ApJ, 497, 759
\bibitem[Fitt \& Alexandern 1993]{fitt-alexander-93} 
   Fitt, A.J., Alexander, P., MNRAS, 261, 445 (1993)
\bibitem[Giacalone and Jokipii 1999]{giacalone99}
   Giacalone, J., Jokipii, R.J., 1999, ApJ, 520, 204
\bibitem[Gressel et al. 2008a]{gressel08a}
   Gressel, O., Ziegler, U., Elstner, D., R
   "udiger, G.
   2008, AN, 329, 61
\bibitem[Gressel et al. 2008b]{gressel08b}
   Gressel, O., Elstner, D., Ziegler, U., R\"udiger, G.
   2008, A\&A, 486L, 35 
\bibitem[Hanasz \& Lesch 1993]{hanasz-lesch-93}
   Hanasz, M., Lesch, H. 1993, A\&A, 278, 561
\bibitem[Hanasz \& Lesch 1997]{hanasz-lesch-97}
   Hanasz, M., Lesch, H. 1997, A\&A, 321, 1007
\bibitem[Hanasz \& Lesch 1998]{hanasz-lesch-98}
   Hanasz, M., Lesch, H. 1998, A\&A, 332, 77
\bibitem[Hanasz \& Lesch 2000]{hanasz-lesch-00}
   Hanasz, M. Lesch, H. 2000, ApJ, 543, 235
\bibitem[Hanasz \& Lesch 2001]{hanasz-lesch-01}
   Hanasz, M. Lesch, H. 2001, Space Sci. Rev., 99, 231.
\bibitem[Hanasz et al. 2002]{hanasz-etal-02}
   Hanasz, M., Otmianowska-Mazur, K., Lesch, H. 2002, A\&A, 386, 347
\bibitem[Hanasz \& Lesch 2003a]{hanasz-lesch-03a}  
   Hanasz, M., Lesch, H. 2003a, A\&A, 404, 389
\bibitem[Hanasz \& Lesch 2003b]{hanasz-lesch-03b}  
   Hanasz, M., Lesch, H. 2003b, A\&A, 412, 331
\bibitem[Hanasz et al. 2004]{hanasz-etal-04}
   Hanasz, M., Kowal, G., Otmianowska-Mazur, K., \& Lesch, H. 2004, ApJ 605, L33
\bibitem[Hanasz et al. 2006]{hanasz-etal-06}
   Hanasz, M., Kowal, G., Otmianowska-Mazur, K., \& Lesch, H.
   2006, AN 327, 469
\bibitem[Hawley, Gammie \& Balbus (1995)]{hawley-etal-95}
   Hawley, J.F., Gammie, C.F., Balbus, S.A. 1995, ApJ, 440, 442
\bibitem[Hessen et al. 2007]{hessen-etal-07}
   Hessen, V., Dettmar, R.-J., Krause, M., Beck R., 2008, arXiv:0801.3542
\bibitem[Jokipii (1999)]{jokipii-99}
   Jokipii, J.R.: 1999, {\em in} J. Franco and A. Carraminana (eds.) 
    {\em Interstellar Turbulence},  Cambridge University Press, 70-78. 
\bibitem[Kowal et al. 2003]{kowal-etal-03}
   Kowal, G., Hanasz, M., Otmianowska-Mazur, K., 2003, A\&A, 404, 533
\bibitem[Kowal, et al. 2005]{kowal-etal-05}
   Kowal, G., Otmianowska-Mazur, K., Hanasz, M., 2005, A\&A, in press
\bibitem[Lesch \& Hanasz 2003]{lesch-hanasz-03}
   Lesch, H., Hanasz, M. 2003, A\&A, 401, 809
\bibitem[Maron \& Blackman 2002]{maron-blackman-02}
   Maron, J., Blackman, E.G. 2002, ApJ, 566, L41
\bibitem[Maron, Cowley \& McWilliams 2004]{maron-etal-04}
   Maron, J., Cowley, S., McWilliams, J. 2004, ApJ, 603, 569
\bibitem[Otmianowska-Mazur, 2003]{otmianowska-03}
   Otmianowska-Mazur, K., 2003, A\&A, 408, 817
\bibitem[Otmianowska-Mazur et al. 2007]{otmianowska-etal-07}
   Otmianowska-Mazur, K.,  Kowal, G., Hanasz, M., 2007, ApJ, 668, 1100
\bibitem[Parker (1992)]{parker-92}
   Parker, E.N. 1992, ApJ, 401, 137
\bibitem[Ryu et al. (2003)]{ryu-etal-03}
  Ryu, D., Kim, J., Hong, S.S., Jones, T.W. 2003, ApJ, 589, 338
\bibitem[Schlickeiser \& Lerche]{schlickeiser-lerche-85}
  Schlickeiser, R., Lerche, I. 1985, A\&A, 151, 151
\bibitem[Snodin et al. (2005)]{snodin-etal-05}
  Snodin, A. P., Brandenburg, A., Mee, A. J., Shukurov, A. 2005,
    astro-ph/0507176
\bibitem[Stone \& Norman 1992a]{stone-norman-92a}
   Stone, J.M., Norman, M.L, 1992a, ApJS, 80, 753
\bibitem[Stone \& Norman 1992b]{stone-norman-92b}
   Stone, J.M., Norman, M.L, 1992b, ApJS, 80, 791
\bibitem[Strong et al. 2007]{strong-etal-07}
   Strong, A.W., Moskalenko, I.V., Ptuskin, V.S.,
   2007, Annual Review of Nuclear and Particle Systems, 57, 285
\bibitem[Tanuma et al. (2003)]{tanuma-etal-03}
   Tanuma, S., Yokoyama, T., Kudoh, T., Shibata, K. 2003, ApJ, 582, 215
\bibitem[Vall\'ee 1995]{vallee-95} 
  Vallee, J.P., A\&A, 296, 819 (1995)

\end{thebibliography}
\end{document}